%% file: main.tex
\documentclass[journal]{IEEEtran}
\IEEEoverridecommandlockouts
\usepackage{cite}
\usepackage{amsmath,amssymb,amsfonts}
\usepackage{algorithmic}
\usepackage{graphicx}
\usepackage{textcomp,subcaption}
\usepackage{xcolor,multicol,multirow}
\usepackage{enumerate,textcomp,array,enumitem}
\usepackage{placeins}
\usepackage{threeparttable}
\usepackage{dblfloatfix}
\usepackage{url}

\usepackage{breakurl}
\usepackage{booktabs}

\usepackage{dblfloatfix}
\usepackage{flushend}

\usepackage{setspace}
\usepackage[running,displaymath]{lineno}
\usepackage{etoolbox} 
\usepackage{makecell}

\newcommand{\tabitem}{~~\llap{\textbullet}~~}
\newcommand*\linenomathpatch[1]{%
  \cspreto{#1}{\linenomath}%
  \cspreto{#1*}{\linenomath}%
  \csappto{end#1}{\endlinenomath}%
  \csappto{end#1*}{\endlinenomath}%
}

\linenomathpatch{equation}
\linenomathpatch{gather}
\linenomathpatch{multline}
\linenomathpatch{align}
\linenomathpatch{alignat}
\linenomathpatch{flalign}

\def\BibTeX{{\rm B\kern-.05em{\sc i\kern-.025em b}\kern-.08em
    T\kern-.1667em\lower.7ex\hbox{E}\kern-.125emX}}

\begin{document}

\title{Interactive Power to Frequency Dynamics Between Grid-Forming Inverters and Synchronous Generators in Power Electronics-Dominated Power Systems}
 	\author{R.~W. Kenyon,~\IEEEmembership{Member,~IEEE},
 		A.~Sajadi,~\IEEEmembership{Senior Member,~IEEE},
 		M.~Bossart,~\IEEEmembership{Student Member,~IEEE},\\
 		A.~Hoke,~\IEEEmembership{Senior Member,~IEEE},
		B.~M.~Hodge*,~\IEEEmembership{Senior Member,~IEEE}%
					
\thanks{R.~W. Kenyon, M. Bossart, and B.~M.~Hodge are with the Electrical Computer and Energy Engineering (ECEE) and the Renewable and Sustainable Energy Institute (RASEI) at the University of Colorado Boulder, Boulder, CO 80309, USA, and the Power Systems Engineering Center, National Renewable Energy Laboratory (NREL), Golden, CO 80401, USA, email: \{richard.kenyonjr,BriMathias.Hodge\}@colorado.edu, \{richard.kenyon,Bri-Mathias.Hodge\}@nrel.gov}
\thanks{A.~Sajadi is with the Renewable and Sustainable Energy Institute (RASEI) at the University of Colorado Boulder, Boulder, CO 80309, USA, email: Amir.Sajadi@colorado.edu}
\thanks{A.~Hoke is with the Power Systems Engineering Center, National Renewable Energy Laboratory (NREL), Golden, CO 80401, USA, email: andy.hoke@nrel.gov}
\thanks{*Corresponding Author: BriMathias.Hodge@colorado.edu}
			}
\maketitle
\vspace{-2cm}
\begin{abstract}
 
With increased attention on grid-forming inverters as a power system stabilizing device during periods of high shares of inverter-based resource operations, there is a present need for a transparent and rigorous investigation of the inverted and direct power to frequency control capabilities, and associated impacts, of these devices on hybrid systems. Here, analysis of the frequency dynamics of the droop controlled grid-forming inverter and the synchronous generator illuminates the inverted active power--frequency relationship and the frequency response order reduction, forming the basis for novel, non-linear frequency droop control approaches. Device-level electromagnetic transient domain simulations corroborate the order-reduction findings, establish that a properly designed DC-side system has a negligible dynamical impact on active power transfer and will not impede frequency regulation, and confirm the frequency response improvement with non-linear control. Simulations of the 9- and 39-bus test systems validate the order reduction and associated decoupling of the nadir and rate of change of frequency in large networks. The primary system oscillatory mode confirms the correlation between high shares of grid-forming inverters and increased mode frequency and damping; a sharp decrease in damping is observed at shares above 80\%, whether by grid-forming device quantity in large networks or rating variations in a small test system. Finally, simulation results on the Hawaiian island of Maui show a trend towards a first order frequency response with a grid-forming inverter, further corroborating the analytic findings and network impacts.


\end{abstract}


\input{sections/1_introduction}

\input{sections/2_power2frequency.tex}
\input{sections/3_math}
\section{Numerical Analysis Results}
\label{sec:num analysis}
This section presents the results of numerical analysis at the device-level and network-level using electromagnetic transient (EMT) simulations. 
The mathematical models presented in Section \ref{sec: converter models} are the foundation for the EMT device implementation in Power System Computer Aided Design (PSCAD)\cite{manitoba_hydro_international_ltd_pscad_nodate}, the software used for all simulations in this work. The devices as implemented are available open source at \cite{kenyon_pypscad_2020}. 
\input{sections/4_numerical_device}
\input{sections/5_numerical_network}
\input{sections/6_maui}

\input{sections/7_discussion}
\input{sections/8_conclusion}

\section*{Acknowledgements}

We wish to thank Jose Daniel Lara and Rodrigo Henriquez Auba with the University of California Berkeley for their insightful discussions and comments. 
 
\bibliographystyle{IEEEtran}

{\scriptsize\bibliography{Wallace_Lib}}

\end{document}

%% file: sections/1_introduction.tex

\section{Introduction}
\label{sec:introduction}

The integration of renewable energy resources into power systems is primarily accomplished with power electronic inverters (i.e., inverter based resources (IBRs)), the majority of which have hitherto employed grid-following (GFL) control strategies that rely on an established voltage and frequency at the point of interconnection to be explicitly tracked with a phase-locked loop (PLL) for operation \cite{kenyon_stability_2020,milano_foundations_2018,kroposki_achieving_2017}. It is well documented that instability can occur for high penetrations of GFL inverters  \cite{nerc_integrating_2017,lin_stability_2017,wang_instability_2018,kenyon_criticality_2023}; operating with 100\% GFL inverters is infeasible. As a result, attention has shifted to grid-forming (GFM) control, where the IBR establishes a local voltage phasor with synchronization objectives \cite{sajadi_synchronization_2022}, such as frequency droop\cite{piagi_autonomous_2006}, and the output frequency of the device is a direct controller action related to delivered active power; a similar relationship is often established for voltage and reactive power \cite{nerc_grid_2021}. The focus of this paper is the interaction of these GFM devices with synchronous generators (SGs) that have an immediate inertial response to active power differentials dictated by physics, as opposed to control design, followed by the slower governor response that manages the active power transfer.

Recent publications on the challenges of frequency stability in low inertia power systems have pointed towards the potential of GFM inverters to mitigate these stability challenges \cite{matevosyan_grid-forming_2019,qoria_deliverable_2018,markovic_understanding_2021,khan_reduced-order_2018,tayyebi_frequency_2020,elkhatib_evaluation_2018,lasseter_grid-forming_2020,lin_research_2020,kenyon_open-source_2021}. The authors of \cite{matevosyan_grid-forming_2019,qoria_deliverable_2018,markovic_understanding_2021,khan_reduced-order_2018} have extensively studied the small-signal stability of power systems with integrated GFMs by developing high-fidelity differential-algebraic models; often, non-zero, minimum SG quantities are declared necessary to preserve system stability, while the opportunity for novel frequency regulation with GFMs is generally neglected. The feasibility of operating bulk power systems with 100\% GFM-based generation has been demonstrated computationally in the electro-magnetic transient (EMT) \cite{tayyebi_frequency_2020,kenyon_open-source_2021}, and phasor \cite{elkhatib_evaluation_2018}, domains, which encourages these devices as a solution to the aforementioned 100\% GFL infeasibility. Tayyebi et al. \cite{tayyebi_frequency_2020} investigated the dynamic interactions between GFMs and SGs, but restricts the approach as GFMs augmenting the SG-driven inertial frequency response. Some studies have begun to identify the damping-like contribution of droop controlled GFM inverters to frequency dynamics \cite{sajadi_synchronization_2022,elkhatib_evaluation_2018,lasseter_grid-forming_2020,tayyebi_frequency_2020}, but once again these view GFM impacts in the context of SG-dominated frequency dynamics. 

\begin{figure}[!htb]
    \centering
    \includegraphics[width=1.0\columnwidth,trim={0 0 0 0},clip]{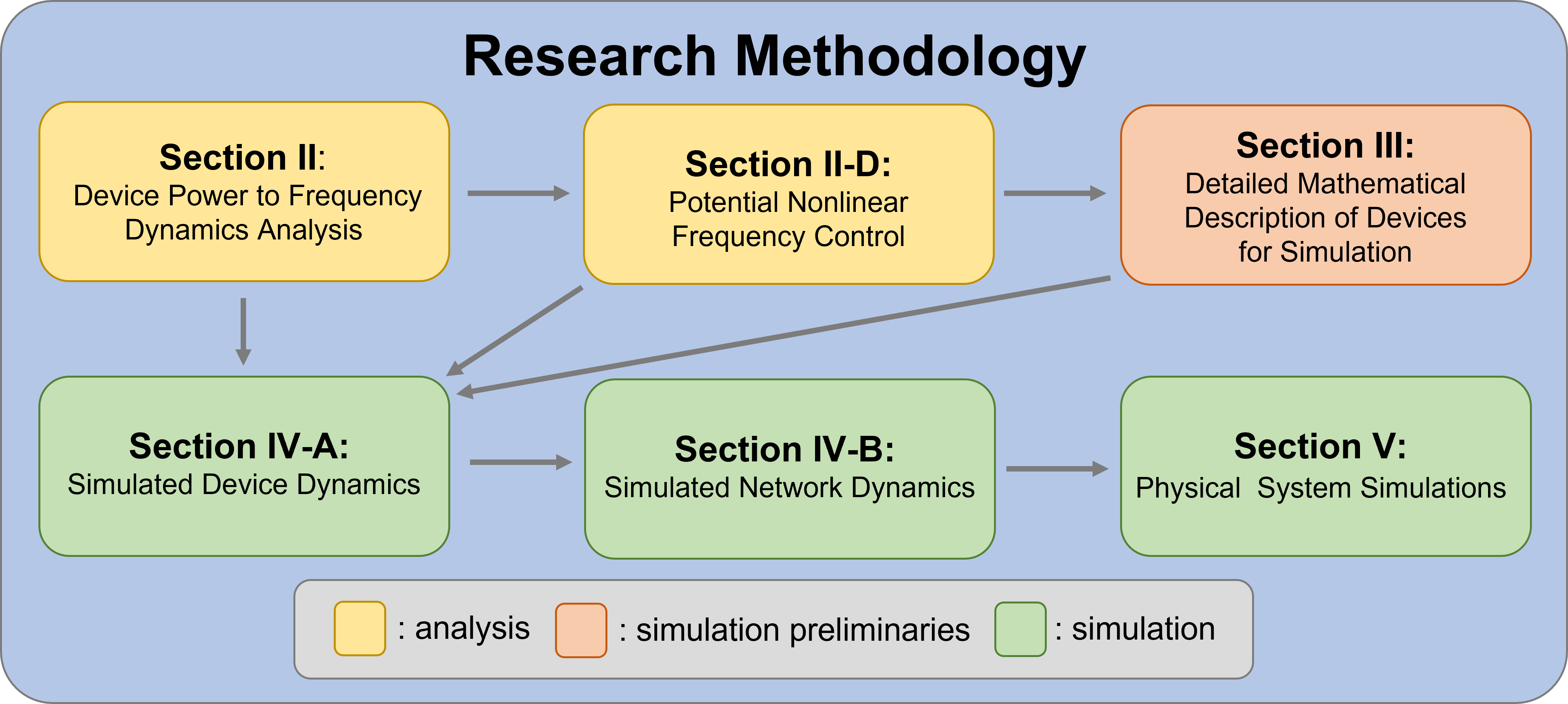}
	\caption{Research methodology employed, with analysis and simulation indicated.}
	\label{fig:research methodology}
\end{figure}

This review of the literature suggests that GFMs are primarily viewed through the lens of SG-driven frequency dynamics, and that there is a present need for an in-depth, side by side investigation of the power to frequency dynamics of these two devices. This paper aims to address this deficit with a fundamental discussion on the inverted operation of GFM inverters (i.e., frequency is calculated and established based on a controller-specified active power delivery relationship), as compared to the reactive primary frequency control of SGs. To this end, the foundational active power transfer dynamics of GFMs and SGs are analyzed to establish the inverted operation and order reduction, which leads to the presentation of novel non-linear frequency control. With the analysis predicated on inverter energy conversion on the order of AC-side delivery, the DC-side dynamics are studied with device level EMT simulations that additionally corroborate the inverted operation and reduced order frequency dynamics. Furthermore, simulations on large test networks, and on a validated model of the Maui island of Hawaii power system, corroborate these order reductions and uncover an unusual oscillatory phenomenon at very low inertia levels. The trajectory of research presented in this work is depicted in Fig. \ref{fig:research methodology}, which shows the general methodology applied and the natural progression of the document.

The three primary contributions of this work are set forth as follows:

\begin{itemize}
    \item \textbf{Contribution 1:} mathematical analysis of the power to frequency dynamics of the SG and GFM devices establishes the inverted active power relationship and order reduction, which supports the presentation of a set of novel non-linear novel frequency droop methods for GFM devices (Section \ref{sec:power to frequency}). 
    \item \textbf{Contribution 2:} it is shown with device-level EMT simulations of full order models of the DC- and AC-side systems (Section \ref{sec: converter models}) that a properly designed DC-side subsystem has a negligible impact on the active power transfer of GFM devices and will not impede frequency regulation (Section \ref{sec:DC side impacts}).
    \item \textbf{Contribution 3:} EMT simulations at the device level confirm the analytic conclusions of contribution 1 (Section \ref{sec:numerical device}), while simulations at the network level with the IEEE 9- and 39-bus test systems (Section \ref{sec:numerical: network}) demonstrate the decoupling of the nadir and rate of change of frequency with GFM devices. This is a departure from the traditional understanding of frequency dynamics \cite{ulbig_impact_2014,tielens_relevance_2016,denholm_inertia_2020}, as a result of the network-wide transition towards first order dynamics. Simulations on a validated model of the Maui power system demonstrate the trajectory to a first order response in a real-world system (Section \ref{sec: maui system}).
\end{itemize}

The remainder of the document now precedes as laid out in the research methodology from Fig. \ref{fig:research methodology}.

%% file: sections/2_power2frequency.tex
\section{Device Power to Frequency Dynamics and Analysis}
\label{sec:power to frequency}

\begin{figure}[h]
    \centering
    \includegraphics[width=0.95\columnwidth,trim={0 0 0 0},clip]{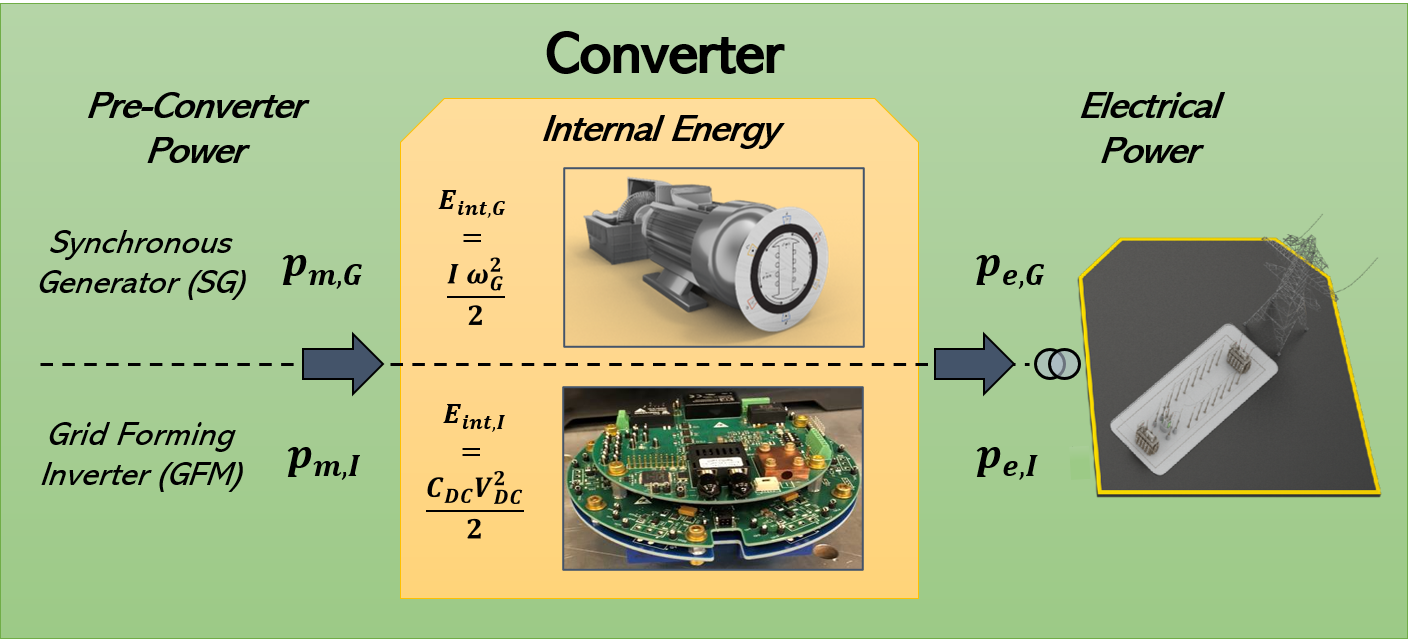}
    \caption{Converter topology showing the relation between the device internal energy ($E_{int})$, pre-converter power ($p_m$), and electrical power ($p_e$). Electronics image from \cite{garrett_development_2019}.}
    \label{fig: power path}
\end{figure}

Each device, i.e., \textit{converter}, converts a pre-converter active power ($p_m$), such as electrochemical energy from a battery for the GFM, or mechanical energy primed by steam, water, or combustion gases for the SG, into AC electrical energy ($p_e$); from here on this is referred to as the \textit{active power through-put process}. Fig. \ref{fig: power path} relates $p_m$ and $p_e$ for each device, where the second subscript (either $_G$ or $_I$) indicates the applicable device (i.e., $_G \rightarrow SG$, and $_I \rightarrow GFM$). Each type of converter stores a quantity of energy, either as kinetic energy for the SG, $E_{int,G} = \frac{1}{2}I\omega_{mech}^2$, with $I$ and $\omega$ being the moment of inertia and shaft rotation rotational speed, respectively, or within the GFM primarily as electrical capacitive storage, $E_{int,I} = \frac{1}{2}C_{DC}V_{DC}^2$, with $C_{DC}$ and $V_{DC}$ being the DC link capacitance and capacitor voltage, respectively. Generally, $E_{int,G} >> E_{int,I}$ \cite{sauer_power_2017,yazdani_voltage-sourced_2010}. Note that in a lossless system, with losses considered negligible here, conservation of energy requires that $\dot{E} \neq 0$ if $p_m \neq p_e$\footnote{dot notation indicates time derivative; $\dot{x} = \tfrac{d}{dt} x$}.

\subsection{Grid Forming Inverter}

A variety of GFM control strategies have been presented in the literature, such as the droop \cite{piagi_autonomous_2006}, multi-loop droop \cite{lasseter_grid-forming_2020}, virtual synchronous machine \cite{beck_virtual_2007}, virtual oscillator control \cite{johnson_synthesizing_2016}, and matching control \cite{arghir_grid-forming_2018}. Generally, they all achieve a similar objective with the construction of a voltage and frequency at the point of interconnection with dynamics associated via a power export relationship. Here, the device of focus is the multi-loop droop control GFM (henceforth, GFM now explicitly means this droop control). The virtual synchronous machine establishes higher order dynamics equitable to SGs by design, and is thus not investigated because the interest here is the contrast of the two devices; the others generally emulate droop control outside of transients. Shown in Fig. \ref{fig: power path}, pre-converter power ($p_{m,I}$) is supplied to the converter via a DC voltage source in the form of electrical or electrochemical energy. A critical assumption in this study is the availability of energy and power via a battery energy storage system\footnote{Other sources of dispatchable power exist for the GFM, such as curtailed photovoltaic output, although here a battery energy storage system operating below its maximum power output is assumed}. The limited power availability condition is outside the scope of this work. Electrical active power ($p_{e,I}$) is fed to the electrical network, with the power to frequency dynamics being directly controlled as the following:
\begin{equation}\label{eq: GFM freq}
    \dot{\delta}_{I} = \omega_{set} + M_p (p^{set} - p_{m,I})\\
\end{equation}
where $\delta_{I}$ is the GFM phase angle, $\omega_n$ is the frequency set point (e.g., 60 Hz in the United States), $M_p$ is the frequency droop value, and $p^{set}$ is the pre-disturbance active power set point.

The distinction between GFM pre-converter power ($p_{m,I}$) and electrical power ($p_{e,I}$) is made to create the comparison analogy; physically, these values are differentiated by a low-pass filter (LPF) in this particular GFM control design. Here, it is assumed that $p_{m,I}$ is readily available within the LPF rise time based on standard energy storage/inverter response times\cite{zhu_impact_2019,kenyon_grid-following_2020}; this assumption is proven valid with an appropriately designed DC-side system in Section \ref{sec:DC side impacts} with EMT simulations. The LPF dynamical relation between $p_{m,I}$ and $p_{e,I}$ is \eqref{eq:GFM power}:
\begin{equation}\label{eq:GFM power}
    \dot{p}_{m,I} = \frac{2\pi\left(p_{e,I} - p_{m,I}\right)}{\tau_{I}}
\end{equation}
where $\tau_{I}$ is the filter time constant. The work in \cite{qoria_deliverable_2018} found a limit of $5\omega_{I,fil} > \omega_n$, which indicates that $\omega_{I,fil}$ should be greater than 75 rad/s in a 60 Hz system, and therefore $\tau_{I} \leq 0.08$ s. 

\subsection{Synchronous Generator}
\label{sec:synch gen}

In modeling of the SG, three distinct components are considered: (1) the machine, (2) the exciter, and (3) the governor/turbine. Here, the \textit{converter} is understood to be the set of all three elements combined. At present, only the machine and governor/turbine dynamics are investigated. The rotational mechanics of the SG are dictated by the swing equation, as shown in \eqref{eq: SG freq} and \eqref{eq: SG rocof}, with the damping component withheld for simplicity:
\begin{align}
    \label{eq: SG freq}\dot{\delta}_{G} &= \omega_G - \omega_s\\
    \label{eq: SG rocof}\dot{\omega}_G &= \frac{1}{M}\left(p_{m,G} - p_{e,G}\right)
\end{align}
where $\delta_{G}$ is the SG phase angle, $\omega_G$ is the SG rotor speed, $\omega_s$ is the synchronous speed (equivalent to $\omega_0$ for a two pole machine), $M = \frac{2H}{\omega_s}$ with $H$ the inertia time constant of the machine, $p_{m,G}$ is the pre-converter mechanical power supplied by the turbine primed by gas, hydro, or steam, and $p_{e,G}$ is the electrical power delivered to the grid. 

The dynamics of $p_{m,G}$ are dictated by the governor, with a variety of different implementations in contemporary practice. However, these dynamics can be broadly described by a dominant filter element, as expressed in \eqref{eq:SG power}:
\begin{equation}\label{eq:SG power}
    \dot{p}_{m,G} = \frac{{R_D}^{-1}\left(\omega_o - \omega_G\right) - \left(p_{m,G} - p_{m,G,o}\right)}{\tau_G}
\end{equation}
where $R_D$ is the droop setting of the device (including a factor of $2\pi$), $p_{m,G,o}$ is the pre-converter power set point, and $\tau_G$ is the governor response time. The value of $\tau_G$ can vary substantially depending on type and model, but is generally not less than 0.5 s \cite{sauer_power_2017}.

\subsection{Device Comparison with Order Reduction}
\label{sec: device comparison}

The active power through-put processes of the GFM (with variables of interest $p_{m,I}$, $\omega_I$, and $p_{m,I}$) and SG (with variables of $p_{m,G}$, $\omega_G$, and $p_{e,G}$) are fundamentally different and these disparate processes heavily influence the frequency response characteristics of the respective devices. In this section the relationships between these processes is analyzed. 

By the control design in \eqref{eq:GFM power}, $p_{m,I}$ has a first order relation to $p_{e,I}$; i.e., $p_{m,I}$ evolves based only on deviations in $p_{e,I}$. Following, $\delta_{I}$ evolves according to \eqref{eq: GFM freq}. Therefore, GFM frequency is a function of the pre-converter power; the device meets the increase in power demand ($p_{e,I}$), and then adjusts the frequency according to the droop relation (recall the assumption of sufficient head room). A GFM does not require local frequency deviations in order for $p_{m,I}$ to evolve. With the SG, $p_{m,G}$ is a function of $\omega_G$ \eqref{eq:SG power}, a relation that is the inverse of the GFM $p_{m,I}$--$\omega_I$ relationship. That is, the governor of the SG modulates $p_{m,G}$ as a function of $\omega_G$, as opposed to the GFM where $\omega_I$ is changed only after the $p_{m,I}$ is matched to the deficit. With droop control, the GFM frequency changes to accomplish power sharing and synchronization which is similar to the Newtonian action of the SG and subsequent governor action; but, the underlying active power through-put power process yields an inverted relationship between frequency and power differentials as compared to the SG. Simply, it can be said that with respect to frequency, whereas the SG is a reactive device, the GFM is a proactive device. 

A further note can be made regarding the $p_m$--$p_e$ relation order of the two devices. With the SG active power through-put process, \eqref{eq: SG rocof} is substituted within the derivative of \eqref{eq:SG power} to produce:
\begin{align}\label{eq:SG 2nd order}
    \ddot{p}_{m,G} &= \frac{{-R_D}^{-1} \dot{\omega}_G - \dot{p}_{m,G}}{\tau_G}\\
    &= \frac{{-(R_DM)}^{-1} \left(p_{m,G} - p_{e,G}\right) - \dot{p}_{m,G}}{\tau_G}
\end{align}
which exhibits that $p_{m,G}$ has a second order relation to $p_{e,G}$, contrasting with the first order dynamics relating $p_{m,I}$ and $p_{e,I}$ of the GFM.

Now, the singular perturbation theory mathematical technique \cite{chow_singular_1990}, which separates varying timescale dynamical systems, is applied to the active power--frequency dynamics analysis of the GFM and SG to discern the respective relationships between $p_{m}$ and $\omega$. Note that both \eqref{eq:GFM power} and \eqref{eq:SG power}, the power conversion dynamical equations for the GFM and SG, respectively, are first order differential equations, with time constants $\tau_{I}$ and $\tau_{G}$, respectively. There is, in general, an order of magnitude of separation between these values; i.e. $\tau_{I} << \tau_{G}$. Therefore, where the time scale of interest is the settling of frequency dynamics associated with the SG, using the foundations of singular perturbation analysis, the follow relation can be established $\dot{p}_{m,I} \approx 0$; thus, according to \eqref{eq:GFM power}, $p_{m,I} \approx p_{e,I}$. 

Due to the relatively slower response of the SG governor, immediately following a disturbance the difference between $p_{m,G,o}$ and $p_{m,G}$ is negligible; i.e., $p_{m,G,o} \approx p_{m,G}$. Applying these approximations, the frequency dynamics relevant before substantial SG governor action of the GFM \eqref{eq: GFM freq} and SG (\eqref{eq: SG freq} and \eqref{eq: SG rocof}) can be reformulated. The approximated GFM frequency dynamics are given in \eqref{eq: GFM freq reduced}:
\begin{equation}\label{eq: GFM freq reduced}
    \begin{split}
    \dot{\delta}_{I} &= M_P (p_{m,I,o} - p_{e,I})\\
    &\propto -p_{e,I}
    \end{split}
\end{equation}
where $\propto$ is the proportional operator. 
The approximated SG dynamics are given in \eqref{eq: SG freq reduced} and \eqref{eq: SG rocof reduced}:
\begin{align}
    \label{eq: SG freq reduced}\dot{\delta}_{G} &= \omega_G - \omega_o\\
\label{eq: SG rocof reduced}
\begin{split}
    \dot{\omega}_G &= \frac{1}{M}\left(p_{m,G,o} - p_{e,G}\right)\\
    &\propto -p_{e,G}
\end{split}
\end{align}
From \eqref{eq: GFM freq reduced}, it can be concluded that $\omega_I$ is proportional to $p_{e,I}$; therefore, considering the first order relation of $p_{e,I}$ and $p_{m,I}$ in \eqref{eq:GFM power}, GFM frequency has a first order relation with pre-converter power, $\dot{\delta}_I \propto p_{m,I}$. From \eqref{eq: SG freq reduced} and \eqref{eq: SG rocof reduced}, with respect to $p_{e,G}$ the frequency dynamics of the SG follow a first order response. Noting the second order relation between $p_{e,G}$ and $p_{m,G}$ in \eqref{eq:SG 2nd order}, it is concluded that the frequency dynamics of the SG have at least a third order relation with respect to pre-converter power within this approximating time frame. 

It is highlighted that the droop gain ($R_D$) parameter for the SG is in the denominator with respect to pre-converter power; \eqref{eq:SG 2nd order}. Therefore, the value of of $R_D$ cannot be zero, while for stability purposes it cannot be near zero due to the resultant rapidity of $p_{m,G}$. Note, $R_D$ is typically held constant (e.g., 5\% in North America) for all frequency responsive devices on a system, for equitable power sharing purposes. For the case of the droop value $M_p$ of the GFM, there is no mathematical limit for the value, while the smaller the value the slower the resultant frequency dynamics, i.e., the ROCOF, due to \eqref{eq: GFM freq reduced}. 

\subsection{Potential for Nonlinear Frequency Control}
\label{sec:potential nonlinear}
Having fully established the inverted relation of active power and frequency of the GFM and determined the timescale reduction of the GFM dynamics, the presentation of potential nonlinear frequency responses for GFM inverters is made. Note that in all of these relations, the primary droop elements are maintained, namely a decrease in frequency for an increase in load, which is necessary to maintain synchronicity \cite{sajadi_synchronization_2022} in AC systems. Consider the basic frequency--active power equation, which is the general form of \eqref{eq: GFM freq}:

\begin{equation}\label{eq:nonlinear}
    \dot{\delta}_I = \omega_{set} - f(p_{m,I})
\end{equation}
where all variables are as previously defined, and $f(p_{m,I})$ is the frequency response function of $p_{m,I}$. In the standard droop formulation, $f(p_{m,I})$ is a linear function, $M_p(p_{m,I})$, where $M_p$ is the droop gain. The correction for initial active power is not explicitly shown. The potential of three other functions is investigated; a quadratic, an exponential (initially presented by the authors in \cite{kenyon_droop-e_2022}), and a power law. The equations are presented in Table \ref{tab:nonlinear functions}, with the chosen constants provided.

\begin{table}[htbp]
    \small
    \centering
    \caption{Nonlinear Functions: Constant values represent those chosen for the comparative simulations}
    {\renewcommand{\arraystretch}{1.0}
    \setlength{\tabcolsep}{.4em}
    \begin{tabular}{c|c|c}
    Type & Function $f(p_{m,I})$ & Constants \\\hline\hline
    Linear & $M_p(p_{m,I})$ & $M_p = 0.05$\\
    Quadratic & $Q_{\alpha} (p_{m,I})^2 + Q_{\beta} (p_{m,I})$ & $Q_{\alpha} = 0.02$, $Q_{\beta}=0.01$\\
    Exponential & $E_{\alpha} e^{(E_{\beta} * p_{m,I})}$ & $E_{\alpha}=0.002$, $E_{\beta}=3.0$\\
    Power & $P_{\alpha} (p_{m,I})^{P_{\beta}}$ & $P_{\alpha}=0.04$, $P_{\beta}=3.0$
    \end{tabular}}
    \label{tab:nonlinear functions}
\end{table}

Figure \ref{fig: Freq Nonlinear} shows the frequency control output for \eqref{eq:nonlinear}, for the four different functions listed in Table \ref{tab:nonlinear functions}. The under frequency load shed (UFLS) potential region is presented for exemplary purposes only; the frequency value of UFLS will vary from  system to system. For a GFM dispatched at 0.1 per unit, the traces show how the frequency output varies based on the delivered power. With the chosen parameters, each of the nonlinear functions will deliver more power with less deviation in frequency, with all three yielding at least 0.2 per unit more active power to the network prior to the UFLS region than the linear droop. The shallower initial tangential droop gains will yield smaller ROCOFs, in accordance to \eqref{eq: GFM freq reduced}. A potential issue with the Power function is the zero valued droop at zero power output; a zero valued droop gain has been shown to be unstable in \cite{kenyon_droop-e_2022}. As $p_{m,I,set}$ changes, the corresponding position on the non-linear curve will change as well; $p_{m,I,set}$ defines where on the curve the 60 Hz crossing occurs. 

\begin{figure}[!htbp]
    \centering
    \includegraphics[width=1.1\columnwidth,trim={0 0 0 40},clip]{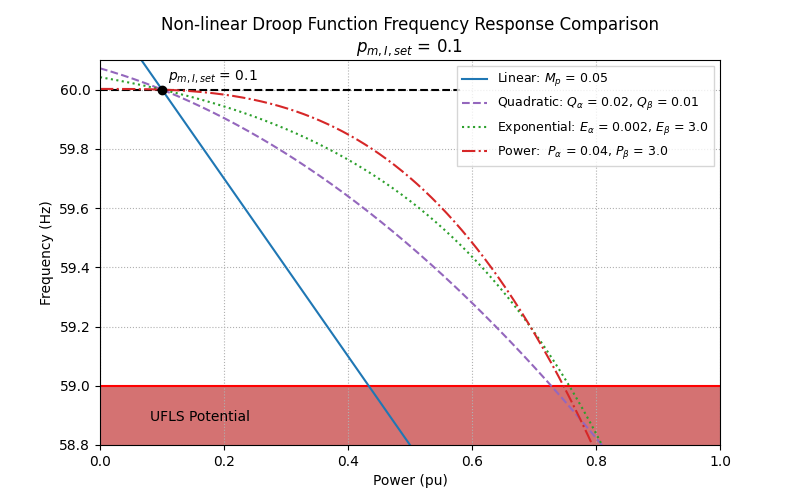}
    \caption{Non-linear droop functions, compared with linear droop. Note, the droop gain changes dynamically, as the GFM power $p_{m,I}$ changes.}
    \label{fig: Nonlinear Curves}
\end{figure}

An obvious deviation in power sharing will occur with these non-linear droop functions. This problem has been solved in \cite{kenyon_droop-e_2022}, where an autonomous power sharing controller was developed that brings the GFM active power output back to an equitable per unit value with other frequency responsive devices on the network. This autonomous controller permits the initial response of the GFM to be whatever the designer desires, while follow on control can achieve power sharing within seconds of the initial deviation. All of these functions are simulated in a small system consisting of a GFM and an SG in Section \ref{sec: nonlinear sims}.

%% file: sections/3_math.tex
\section{Detailed Mathematical Description of Converters for Simulation}
\label{sec: converter models}

\begin{figure*}[h]
    \centering   \includegraphics[width=1.85\columnwidth,trim={0 0 0 0},clip]{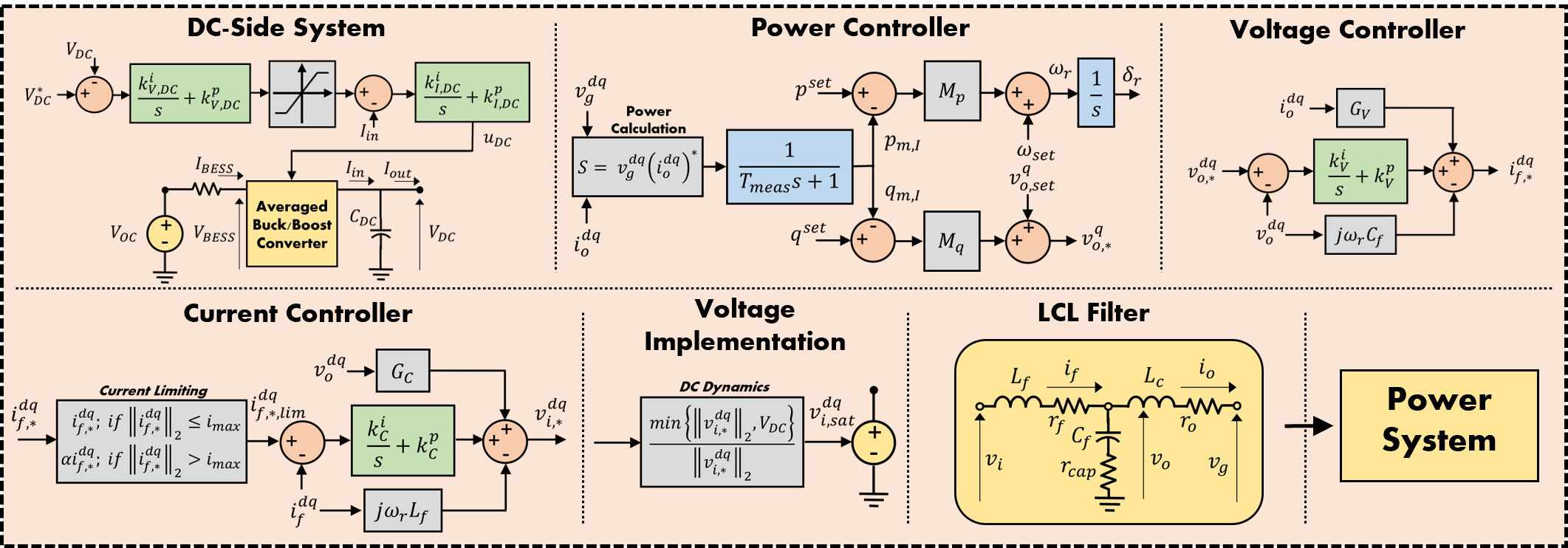}
    \caption{Control scheme of multi-loop droop grid-forming inverter. All elements shown are considered part of the converter, as displayed in Fig. \ref{fig: power path} -- some interconnecting signals are not explicitly shown.}
    \label{fig:GFM Control}
\end{figure*}

The following discusses the mathematical models constructed for the representation of the GFM and SG devices in the EMT simulations.

\subsection{Grid Forming Inverters}
The control scheme of the GFM inverter is shown in Fig. \ref{fig:GFM Control}. High frequency pulse width modulation and switching is not modelled, with averaged voltage signals being provided to ideal voltage sources as a standard power electronics approximation \cite{erickson_fundamentals_2007}. In this control, an LC output filter meets with a coupling inductor to create the LCL topology shown in Fig. \ref{fig:GFM Control}. The filter inductor current ($i_f$) and filter capacitor voltage ($v_o$) are regulated by proportional-integral (PI) controllers operated in the direct-quadrature (\textit{dq}) synchronous reference frame. That is, $\bar{x}_{dq} = T_{dq}\bar{x}_{abc}$, where $\bar{x}_{abc}$ is a three dimensional column vector representing the sinusoidal three phase waveforms, $\bar{x}_{dq}$ is a two dimensional column vector representing the non-sinusoidal \textit{d} and \textit{q} axis waveforms, and $T_{dq}$ is the  $2\times3$ Park transformation matrix \cite{park_two-reaction_1929}.

The instantaneous active ($p_{e,I}$) and reactive ($q_{e,I}$) powers in the $dq$ frame are calculated as $p_{e,I}=\Re{(v^{dq}_gi_o^{dq,*})}$\footnote{$^*$ denotes the complex conjugate} and $q_{e,I}=\Im{(v^{dq}_gi_o^{dq,*})}$, where $v^{dq}_g$ and $i^{dq}_o$ are the $dq$ frame grid voltages and coupling filter currents, respectively, represented in the complex domain (i.e., $\alpha + j\beta$). The instantaneous powers $p_{e,I}$ and $q_{e,I}$ are passed through a LPF with cutoff frequency $\omega_{I,fil}$, as shown in \eqref{eq: p filter} (reformulated from \eqref{eq:GFM power}, where $\omega_{I,fil} = 2\pi/\tau_{I}$) and \eqref{eq: q filter}:
\begin{align}
    \label{eq: p filter}\dot{p}_{m,I} &= \omega_{I,fil}(p_{e,I}-p_{m,I})\\
    \label{eq: q filter}\dot{q}_{m,I} &= \omega_{I,fil}(q_{e,I}-q_{m,I})
\end{align}

The currents across the filter resistor ($R_f$) and inductor ($L_f$), $i_f^{dq}$ are regulated with proportional integral (PI) controllers (depicted by the Current Controller block in Fig. \ref{fig:GFM Control}):
\begin{align}
    \label{eq: current controller int}\dot{\gamma}^{dq} &= i_f^{dq} - i^{dq}_{f,*}\\
    \label{eq: current controller} v_{i,*}^{dq} &= k_{C}^{i} \gamma^{dq} + k_{C}^p \dot{\gamma}^{dq} - j\omega_I L_f i_f^{dq}+G_{C}v_o^{dq}
\end{align}
where $\gamma^{dq}$ are integrator error states (not mapped in the imaginary plane), $k_C^i$ and $k_C^p$ are the integral and proportional gains, respectively, $\omega_I$ is the radian frequency from the droop relation and $G_{C}$ is the voltage feed forward gain. The controller gains are tuned to cancel the inductor pole. The response time of the inner current loop is on the order of 1 millisecond. The capacitor ($C_f$) voltage is also regulated with a PI controller, as depicted by \eqref{eq: VC int} and \eqref{eq: VC PI} (depicted by the Voltage Controller block in Fig. \ref{fig:GFM Control}):
\begin{align}
    \label{eq: VC int}\dot{\xi}^{dq} &= v_o^{dq} - v^{dq}_{o,*}\\
    \label{eq: VC PI}i_{f,*}^{dq} &= k_V^{i} \xi^{dq} + k_V^p \dot{\xi}^{dq} - j\omega_I C_f v_o^{dq}+G_{V}i_o^{dq}
\end{align}
where $\xi^{dq}$ are the integrator error states (not mapped in the imaginary plane), $k_V^i$ and $k_V^p$ are the integral and proportional gains, respectively, and $G_V$ is the current feed forward gain. The response time of the voltage controller is on the order of 10 milliseconds. The fundamental governing equations of the multi-loop droop grid forming inverter are given by the previously discussed \eqref{eq: GFM freq} and \eqref{eq: GFM volt}, which are represented in the Power Controller block in Fig. \ref{fig:GFM Control}:
\begin{equation}\label{eq: GFM volt}
v_{o,*}^q = v_{o,set}^q + M_q (q^{set} - q_{m,I})
\end{equation}
where $q^{set}$ is the pre-disturbance power set points, $p_{m,I}$ and $q_{m,I}$ are as defined in \eqref{eq: p filter} and \eqref{eq: q filter}, $v_{o,*}^q$ is the set point for the voltage controller ($v_{o,*}^d = 0$), $v_{o,set}^q$ is the pre-disturbance steady state voltage, and $M_q$ is the voltage droop gain. 

A zeroth order model with a voltage source and resistor is used to represent the battery as shown the DC-Side System panel of Fig. \ref{fig:GFM Control}; higher order battery models, with series RL/RC branches, had a negligible impact on the battery response\cite{roberts_grid-coupled_2020}. The governing equation of the battery output voltage is therefore \eqref{eq:Bess}:
\begin{equation}\label{eq:Bess}
    v_{BESS} = v_{OC} - i_{BESS}R_{BESS}
\end{equation}
where $V_{BESS}$ is the output voltage, $R_{BESS}$ is the battery resistance, $i_{BESS}$ is the delivered current, and $V_{OC}$ is the battery characteristic open circuit voltage.

The voltage, $V_{BESS}$ is passed to a boost converter to regulate the voltage, $v_{DC}$ across the capacitor $C_{DC}$, which forms the voltage source for the inverter. Figure \ref{fig:GFM Control} depicts the boost converter and battery control block diagram. The dynamical equations describing the cascading PI controllers are:
\begin{align}
    \label{eq:DC V Int}\dot{\chi} &= v^*_{DC} - v_{DC}\\
    \label{eq:DC V Control}i_{ref} &= k_{V,DC}^i\chi + k_{V,DC}^p\dot{\chi}\\
    \label{eq:DC I Int}\dot{\zeta} &= i_{ref} - i_{in}\\
    \label{eq:DC I Control}u &= k_{i,DC}^i\zeta + k_{i,DC}^p\dot{\zeta}
\end{align}
where $\chi$, $k_{V,DC}^i$, and $k_{V,DC}^p$ are the integrator state, integral gain, and proportional gain of the DC voltage controller, respectively. $i_{ref}$ is limited to $||i_{ref}|| \leq i_{rate}$. $\zeta$, $k_{i,DC}^i$, and $k_{i,DC}^i$ are the integrator state, integral gain, and proportional gain of the DC current controller, respectively. The output from the DC voltage controller, is limited to a range of values between $[0.05,0.90]$. The inverter operation is impeded by the DC side dynamics when the 2-norm of the commanded voltages ($||v_{i,*}^{dq}||_2$) exceeds the ($v_{DC}$), indicating saturation. This is modelled by scaling the commanded voltages based on the minimum function between the 2-norm of $v_{i,*}^{dq}$ and $v_{DC}$, as in \eqref{eq: DC sat} \cite{roberts_grid-coupled_2020,pico_transient_2019}:
\begin{align}
    \label{eq: DC sat}v_{i,sat}^{dq} &= \left(\frac{min\{||v_{i,*}^{dq}||_2,v_{DC}\}}{||v_{i,*}^{dq}||_2}\right)v_{i,*}^{dq}\\
    \label{eq: 2 norm}||v_{i,*}^{dq}||_2 &= \sqrt{(v_{i,*}^{d})^2 + (v_{i,*}^{q})^2}
\end{align}
where $v_{i,sat}^{dq}$ is the implemented voltage command, and $v_{i,*}^{d}$ and $v_{i,*}^{q}$ are the $d$ and $q$ axis components of the voltage commands, respectively. Note that $v_{i,sat}^{dq} = v_{i,*}^{dq}$ when $v_{DC}$ exceeds $||v_{i,*}^{dq}||_2$. The coupling variable between the DC side control and the AC side implementation is $v_{i,sat}^{dq}$, which is applied to the voltage source $v_i$ in the Voltage Implementation panel of Fig. \ref{fig:GFM Control}).

\subsection{Synchronous Generator}
\label{sec: SG}

\begin{figure}[h]
    \centering
    \includegraphics[width=0.95\columnwidth,trim={0 0 0 0},clip]{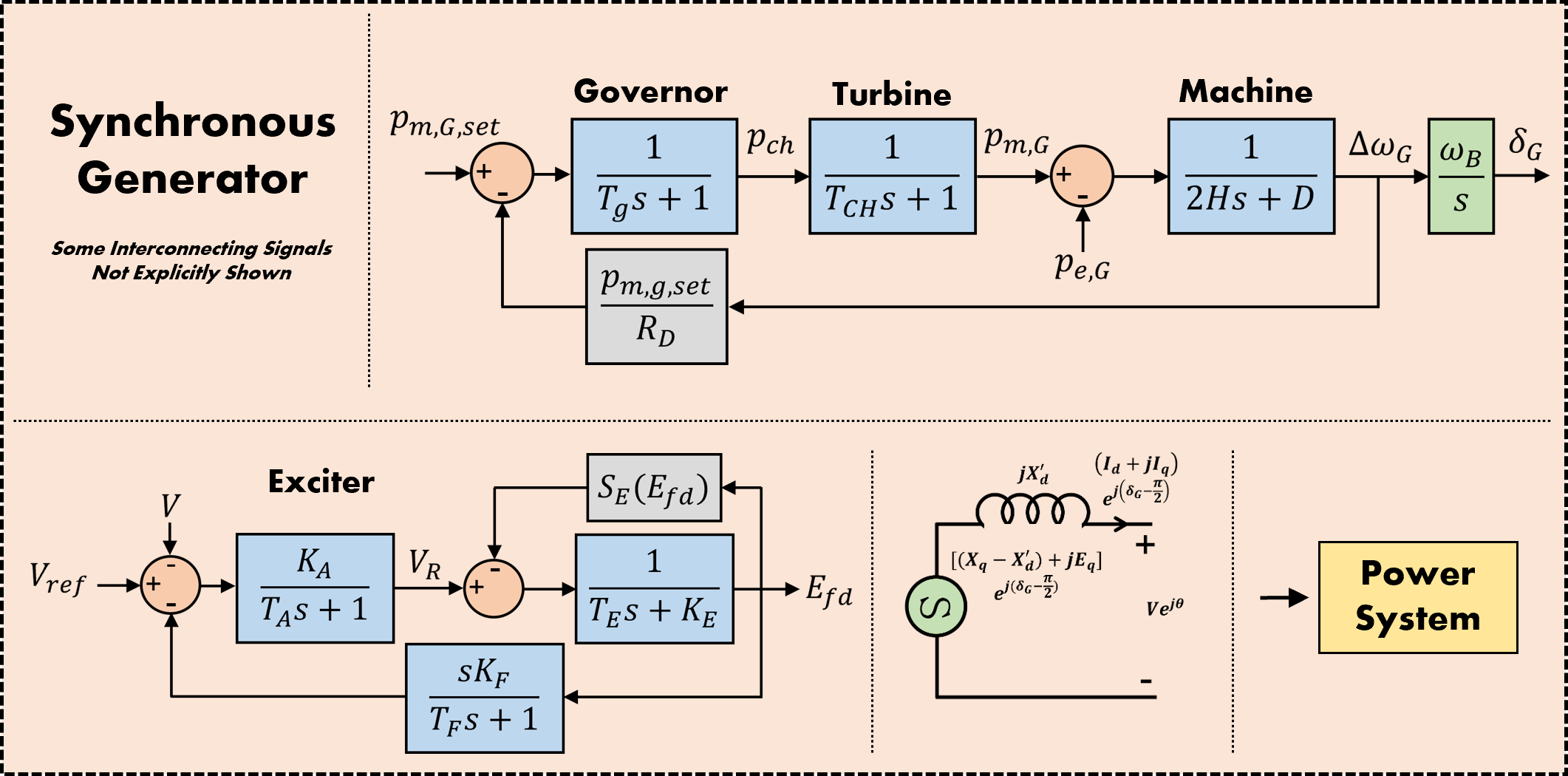}
    \caption{Synchronous generator control implementation.}
    \label{fig:SG control}
\end{figure}

The SG is a well documented device; see \cite{machowski_power_2008,sauer_power_2017} for in-depth discussions. In the EMT modeling of the SG, all three distinct components are modeled: (1) the machine, (2) the exciter, and (3) the governor/turbine. The overarching control implementation for the EMT simulations is given in Fig. \ref{fig:SG control}, while it is noted that many limiting functionalities are not shown.

The fundamental machine model is a set of differential current elements in the \textit{dq} frame expressed as a function of flux linkages and applied voltages\cite{manitoba_hydro_international_ltd_pscad_nodate}. The machines in this study are established with two \textit{q} axis damper windings, consistent with the round-rotor implementation. The swing equation is depicted once again with damping for transparency:
\begin{align}
    \dot{\delta}_{G} &= \omega_G - \omega_s\\
    \label{eq: SG rocof damping}\dot{\omega}_G &= \frac{1}{M}\left(p_{m,G} - p_{e,G} - D\dot{\delta}_{G}\right)
\end{align}
where $D$ is the damper winding component.

The exciter is based on the IEEE Type-1 model. The saturation function is an exponential of the form: $S_E(E_{fd}) = \gamma e^{\epsilon E_{fd}}$. No derivative gains or derivative feedback are modelled. The governor/turbine model used here is a simple representation that acts on speed (frequency) errors \cite{sauer_power_2017}. The model consists of two cascaded low pass filters and limiters associated with output limits. 

\subsection{Simulation Statistics}
When applicable, the system equivalent inertia ($H$) is defined based on the relative ratings of the devices as \eqref{eq:aggregate inertia}:
\begin{equation}\label{eq:aggregate inertia}
    H = \frac{\sum_{i=1}^n H_i S_{B,i}}{\sum_{i=1}^nS_{B,i}}
\end{equation}
where $H_i$ is the inertia rating (in $s$) of device $i$, $S_{B,i}$ is the MVA rating of device $i$, and $n$ is the number of devices. For a GFM, $H=0$.

Unless otherwise specified, the frequency of the system under inspection is the weighted frequency of all devices, which is calculated according to \begin{equation}\label{eq:measurefrequency}
    f(t) = \frac{\sum_{i=1}^n (MVA_i*f_i(t))}{\sum_{i=1}^n MVA_i}
\end{equation}
where $f_i(t)$ is the frequency of device $i$ at time $t$, $MVA_i$ is the device $i$ rating, and $n$ is the number of devices. This weighted frequency is used to determine the ROCOF and nadir values. The ROCOF is defined as:
\begin{equation}\label{eq:ROCOF} 
\dot{f}(t) = \frac{f(t + T_{R}) - f(t)}{T_{R}}
\end{equation}
where $f$ is the frequency, and $T_{R}$ is the size, in seconds, of the sliding averaging window. A $T_{R} = 100\hspace{0.1cm} ms$ window is used, in accordance with \cite{inverter-based_resource_performance_task_force_fast_2020}. The nadir is defined as $min\{f(t)\}$, where $t>t_{perturbation}$; i.e., the lowest frequency post disturbance. The matrix pencil method\cite{sarkar_using_1995
} is used to calculate the frequency ($\omega_k$) and decay ($\lambda_k$) of the modes. The damping factor is then $\label{eq: damping ratio} \zeta_k = \frac{\lambda_k}{\sqrt{\lambda_k^2 + \omega_k^2}}$ where $k$ corresponds with the dominant mode.

%% file: sections/4_numerical_device.tex
\subsection{Device-Level Dynamics}
\label{sec:numerical device}
The individual device responses are presented followed by the direct interaction of the two devices as analyzed on a three bus system, demonstrating the interrelated dynamics. The device-level analysis is concluded with an investigation into the impacts of the DC-side dynamical system, and simulations with the non-linear frequency control functions.

\subsubsection{Device Step Response}
\label{sec: step response}

To verify the analytical findings and conclusions of Section \ref{sec:power to frequency}, the frequency response of the SG and GFM following a load perturbation is simulated. For this analysis, each device is isolated, dispatched at 0.5 per unit (pu) with a constant power load connected directly to the terminals. 

\begin{figure}[h]	
	\centering
	\begin{subfigure}[t]{1.6in}
		\centering
		\includegraphics[trim=5 5 40 39,clip,width=1\textwidth]{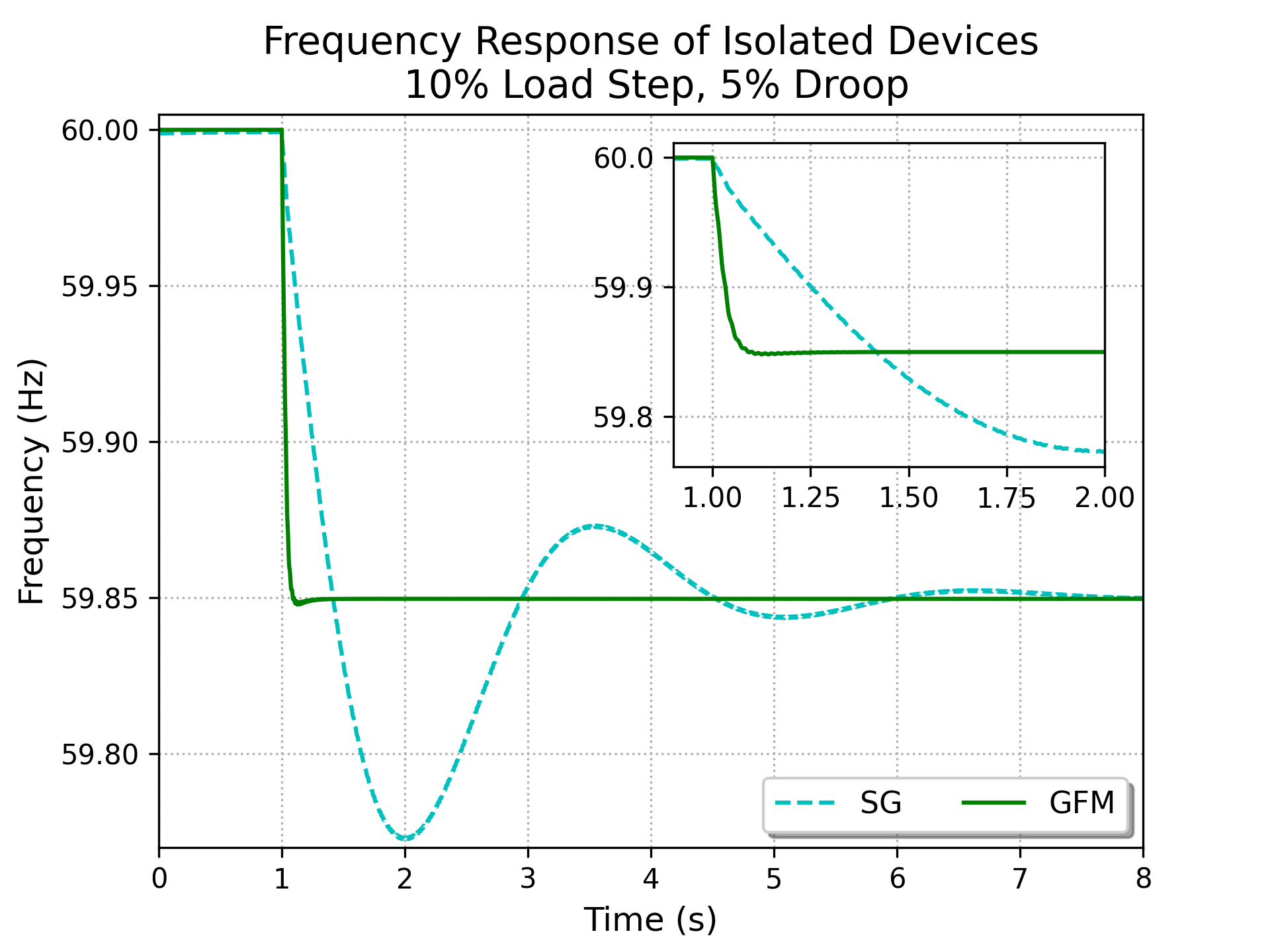}
		\caption{Frequency response.}\label{fig: frequency load step}	
	\end{subfigure}
	\quad
	\begin{subfigure}[t]{1.6in}
		\centering
		\includegraphics[trim=5 5 40 39,clip, width=1\textwidth]{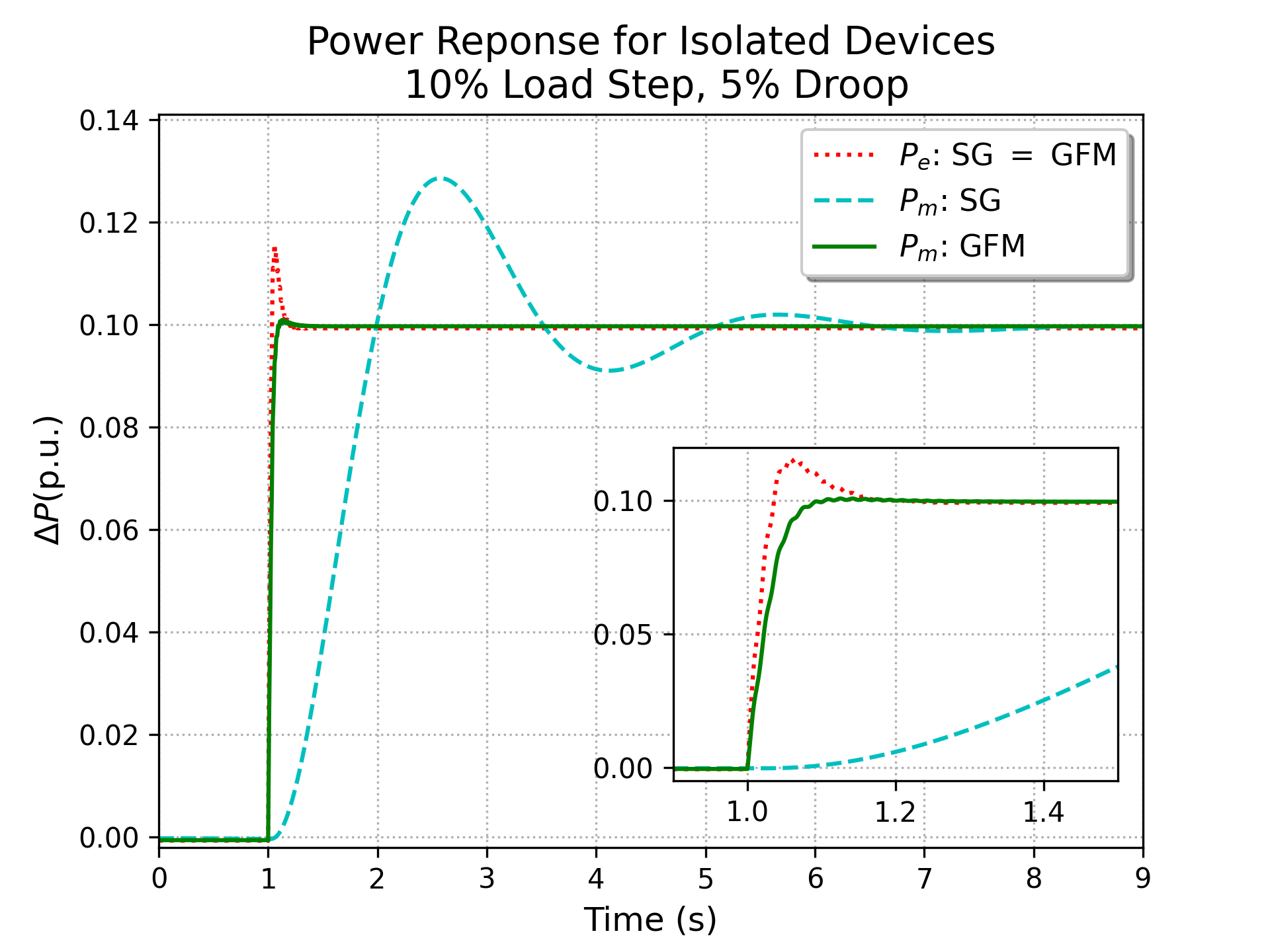}
		\caption{Power response.}\label{fig: power load step}
	\end{subfigure}
	\caption{Frequency and power responses of isolated GFM and SG devices following a 0.1 pu load step.}\label{fig: Device Step Response}
\vspace{-1em}	
\end{figure}

The frequency and power step response of each device for a 0.1 pu load step are presented in Fig. \ref{fig: Device Step Response}. The SG frequency trace in Fig. \ref{fig: frequency load step} shows the second order negative step response following the load step with overshoot and subsequent damped oscillations that settle to the droop determined steady state. The GFM frequency trace follows a first order response; there is no overshoot and a far smaller response time as compared to the SG. This result is due in part to the faster response of the GFM, but primarily a dynamical system order reduction. A summary of the nadir, settling frequency, and ROCOF are presented in Table \ref{tab:Single Device Results}. Note the inverse proportionality between inertia and ROCOF and the correlation between a lower nadir and reduced inertia for SGs, contrasted with the larger ROCOF of the GFM device, but a resultant higher nadir. 

\begin{table}[!htbp]
    \small
    \centering
    \caption{Single Device Step Response Results}
    \begin{tabular}{c|c|c|c}
     & ROCOF & Nadir & Settling\\
    Device &(Hz/s) & (Hz) & Frequency (Hz)\\\hline\hline
    SG ($H=4s$) & 0.48 & 59.77 & 59.85\\
    SG ($H=3s$) & 0.63 & 59.73 & 59.85\\
    SG ($H=2s$) & 0.95 & 59.67 & 59.85\\
    SG ($H=1s$) & 1.90 & 59.52 & 59.85\\
    GFM & 1.50 & 59.85 & 59.85\\
    \end{tabular}
    \label{tab:Single Device Results}
    \vspace{-1em}
\end{table}

The $p_e$ and $p_m$ responses of each device are presented in Fig. \ref{fig: power load step}. Note that the $p_e$ response is identical for each device; the electrical power is primarily determined by network changes because the output voltage is held nearly constant. The first order relation of $p_{m,I}$ and $p_{e,G}$ is evident in the GFM traces, which manifest at a faster rate than the higher order SG response, corroborating the timescale separation impetus for the singular perturbation application. The $p_{m,G}$ follows a second order response complete with overshoot, oscillations, initial acceleration, and inflection change.

\subsubsection{Two-generator System}
\label{sec:three bus comp}

A simple two generator test system was constructed with only a single GFM (bus 1), a single SG (bus 3), and a single load (bus 2), connected by equal impedance lines ($X_T = 0.025 + j0.125$). 
A set of 11 simulations were performed, wherein the aggregate rating of the devices was maintained at 30 MVA to sweep the system equivalent inertia ($H$) space from 4.0 -- 0.0 s. Table \ref{tab:3 Bus Results} provides these device ratings and respective inertia. These rating ratios will match the aggregate inertia ratings for the 39-bus system simulation scenarios in Section \ref{sec: 39 bus}. The devices are dispatched at 0.5 pu relative to the device rating, and the perturbation is a 0.1 pu load step relative to the 30 MVA total of the system.

\begin{figure}[htbp]
    \centering
    \includegraphics[width=0.8\columnwidth,trim={0 0 0 0},clip]{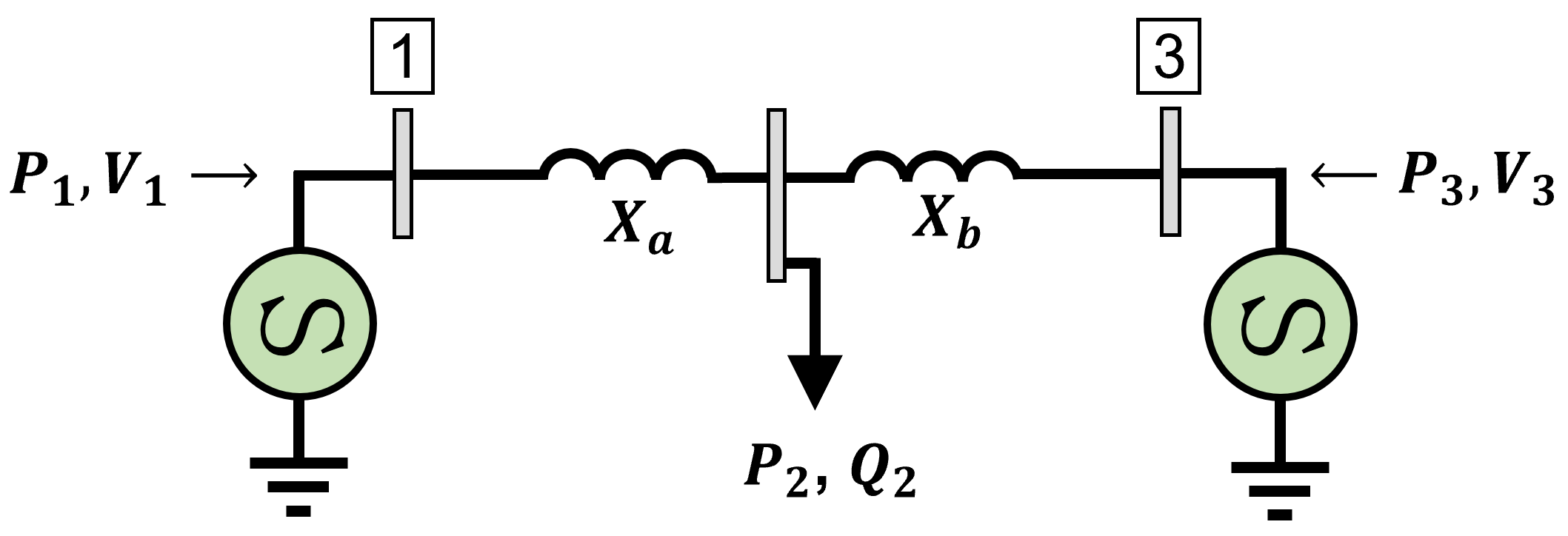}
    \caption{Three bus test system used for the rating sweep, DC-level analysis, and nonlinear frequency control demonstration.}
    \label{fig: 3 bus system}
\end{figure}

\begin{table}[htbp]
    \small
    \centering
    \caption{Two Generator System Results}
    \setlength\tabcolsep{2pt}   
    \begin{tabular}{c|c|c|c|c}
    SG Rating & GFM Rating & Inertia & SG ROCOF & SG Nadir\\
    (MVA) & (MVA) & (s) & (Hz/s) & (Hz) \\\hline\hline
    30 & n/a & 4.0 & 0.426 & 59.687\\
    27 & 3 & 3.6 & 0.489 & 59.701\\
    24 & 6 & 3.2 & 0.528 & 59.725\\
    21 & 9 & 2.8 & 0. 582 & 59.740\\
    18 & 12 & 2.4 & 0.645 & 59.754\\
    15 & 15 & 2.0 & 0.721 & 59.765\\
    12 & 18 & 1.6 & 0.813 & 59.775\\
    9 & 21 & 1.2 & 0.934 & 59.785\\
    6 & 24 & 0.8 & 1.102 & 59.791\\
    3 & 27 & 0.4 & 1.361 & 59.767\\
    n/a & 30 & 0.0 & 1.390 (GFM) & 59.842 (GFM)
    \end{tabular}
    \label{tab:3 Bus Results}
    \vspace{-1em}
\end{table}

\begin{figure}[htbp]
    \centering
    \includegraphics[width=0.8\columnwidth,trim={0 0 0 40},clip]{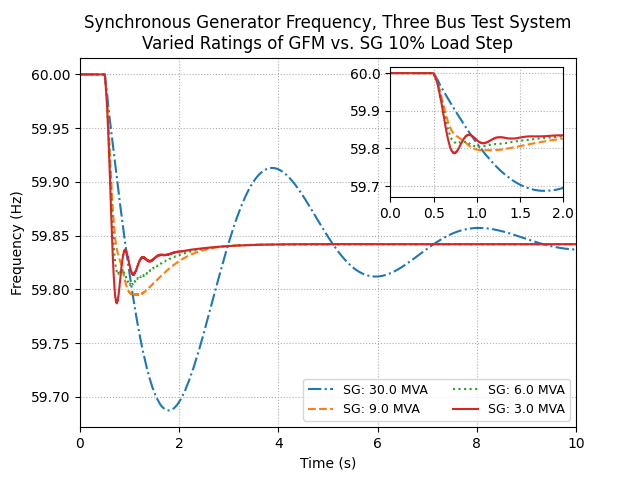}
    \caption{Frequency response of the synchronous generator (SG) from the two generator test system with a GFM, SG, and load.
    }
    \label{fig: 3 bus SG freq}
    \vspace{-1em}
\end{figure}

The frequency results presented in Table \ref{tab:3 Bus Results} are for the SG alone (except for the final row), to highlight the impact of the GFM on the SG operation. There is a clear shift with a growing GFM ratio to an increase in ROCOF, accompanied by a rising nadir. Figure \ref{fig: 3 bus SG freq} shows the SG frequency for a few of these cases, where it is evident that substantial damping is afforded to the SG response, until the SG represents only a small portion of  the system machine rating (i.e., $< 20\%$). When the SG rating is 3 MVA, and the GFM is rated at 27 MVA, a high frequency (2.9 Hz) oscillatory mode is present that dampens out within a few cycles. This oscillatory mode is present only as rapid variations in $p_{e,G}$,; $p_{m,G}$ does not change within the time frame of this oscillatory response. This mode is much faster than the 0.23 Hz oscillatory mode dominating the SG only simulation (SG: 30 MVA). The primary takeaway is that for low levels of SGs as compared to GFMs, with respect to total rating, high frequency oscillatory modes are excited in the SG rotors. The anticipated cause for these modes is the result of rapid power flow differentials generated on the network by the rapid GFM frequency change. This will incur a substantial $p_{e,G} - p_{m,G}$ differential on the SG, that when matched with the low inertia, would result in these higher frequency oscillatory modes. Similar tests with the AC1A exciter yielded no difference in these simulation results, with the response of the exciter being substantially faster than the active power transfer dynamics of the GFM and SG. Different governor models such as GGOV1, TGOV1, and GAST were simulated, but with bandwidths outside of the oscillatory mode, no impact was expected or observed. Further investigation into the root cause of these oscillatory modes is a prime topic for future research.

\subsubsection{Impact of DC-Side Dynamics}
\label{sec:DC side impacts}
In Section \ref{sec:power to frequency}, it was assumed that $p_{m,I}$ is immediately available to the GFM upon changes in $p_{e,I}$. This assumption is based on the suitable operation of the DC-side system, which is predicated on the appropriate device size selection and boost converter operation. This system is now analyzed to confirm the assumption. To determine an appropriate capacity of the DC-link capacitor ($C_{DC}$), and a suitable DC-link voltage across the capacitor ($V_{DC}$), a computational sensitivity analysis was performed. The capacitor sizes investigated represent either 0.1, or 1 cycle's worth of energy contained at the rated inverter power and voltage. That is, the capacitor rating is determined according to \eqref{eq: C rating}:
\begin{equation}\label{eq: C rating}
    C_{DC} = \frac{2 S_{base} \alpha}{v_{DC}^2}
\end{equation}
where $S_{base}$ is the device rating and $\alpha$ represents the time duration based on the cycles of rated power. At a minimum, the $V_{DC}$ must span the positive and negative peaks of the inverter output. Namely, $v_{DC} > V_{base,l-l}\sqrt{8/3}$. Factoring in the anticipated operation of the inverter above rated voltage, in this case it is assumed 1.05 pu, the target $V_{DC}$ can be represented as in \eqref{eq: DC Set}:
\begin{equation}\label{eq: DC Set}
    v_{DC,set} = 1.05 \sqrt{8/3} V_{base,l-l}\Delta V
\end{equation}
where $\Delta V$ is a parameter that represents the setpoint overage above the requisite voltage. The parameter sweep is an executed 0.5 pu load step, with an initial dispatch at 0.05 pu, on an isolated GFM inverter for six different value combinations of $C_{DC}$ and $\Delta V$. The pu power differential between the no DC dynamics ($p_{noDC}$) and the resultant power with the specific parameters ($p_{DC}$) is integrated (discretely) across the length of the simulation, {$E_{diff}=\sum (p_{no DC} - p_{DC})\Delta t$}, to quantify the impacts of the DC side dynamics. These results are shown in Table \ref{tab: DC compare}, where it is evident that for both voltage buffers of $\Delta V = 5\%$ and $\Delta V = 10\%$, the differential is very small; it is effectively zero when the capacitor is rated for one electrical cycle and $\Delta V = 10\%$. In simulations where $E_{diff} \neq 0$, the difference persists for less than 200 milliseconds. Subsequently, the DC side values of $C_{DC} = 1$ cycle and $\Delta V = 10\%$ are selected to support the immediate availability of pre-converter power assumed in the analysis of Section \ref{sec:power to frequency}. Having concluded on this appropriate assumption, DC side dynamics were disabled in the following simulations to improve simulation time.

\begin{table}[!ht]
    \renewcommand{\arraystretch}{1.2}
    \setlength{\tabcolsep}{1em}
    \centering
    \caption{DC Side Energy Difference - The resultant $E_{diff}$ values are expressed in per unit * seconds}
    \small
    \label{tab: DC compare}
    \begin{tabular}{cc||c|c}
        \multicolumn{2}{c}{\multirow{2}{*}{~}} & \multicolumn{2}{c}{{Cycles of Capacitive Energy}}\\
        \multicolumn{2}{c||}{}& \phantom{0000}0.1\phantom{0000}& 1\\\cline{2-4}\noalign{\vskip\doublerulesep
         \vskip-\arrayrulewidth}\cline{2-4}
        \multirow{3}{*}{\rotatebox{90}{$\Delta V$ (\%)}} & 0 & $0.198$ & $0.159$ \\\cline{2-4}
        & 5 & $0.031$ & $0.003$\\\cline{2-4}
        & 10 & $0.0001$ & $0.0$\\
    \end{tabular}
    \vspace{-1em}
\end{table}

\subsubsection{Nonlinear Frequency Control Simulations}
\label{sec: nonlinear sims}

\begin{figure}[!htbp]
    \centering
    \includegraphics[width=0.95\columnwidth,trim={0 0 0 39},clip]{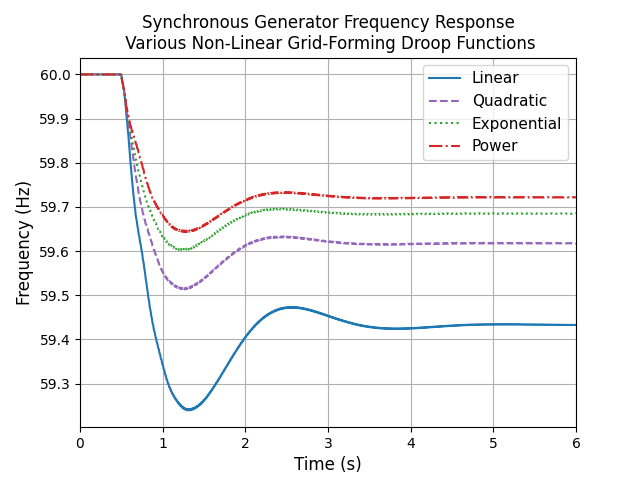}
    \caption{Frequency of the SG on the 3-bus system for various frequency droop functions of the GFM.}
    \label{fig: Freq Nonlinear}
\end{figure}

Test results with the non-linear frequency functions from Section \ref{sec:potential nonlinear} are presented here, with the function parameters as presented in Table \ref{tab:nonlinear functions}. The three-bus system from Fig. \ref{fig: 3 bus system} is used, where the SG is rated at 30 MVA and dispatched at 0.5 per unit, and the GFM is rated at 15 MVA and dispatched at 0.1 per unit. A 2.5 MW load step at bus 3 is executed. The frequency responses presented in Fig. \ref{fig: Freq Nonlinear}, with statistics in Table \ref{tab: non-linear droop response}, are derived from the rotational velocity of the SG. All three non-linear functions present improved frequency dynamics, both in the nadir and ROCOF, as compared to the linear function, while the primary oscillatory mode exhibits better damping with all of the non-linear responses.

\begin{table}[htbp]
    \small
    \centering
    \caption{Non-linear Droop }
    \begin{tabular}{c|c|c|c|c}
     & ROCOF & Nadir & \multicolumn{2}{c}{$\Delta P_{tot}$ (p.u.)}\\
    Device &(Hz/s) & (Hz) & \multicolumn{1}{c}{SG} & \multicolumn{1}{c}{GFM}\\\hline\hline
    Linear & 2.14 & 59.23 & 0.056 & 0.056\\
    Quadratic & 1.39 & 59.51 & 0.037 & 0.092\\
    Exponential & 1.17 & 59.6 & 0.030 & 0.105\\
    Power & 1.06 & 59.64 & 0.027 & 0.112\\
    \end{tabular}
    \label{tab: non-linear droop response}
\end{table}

\begin{figure}[!htbp]
    \centering
    \includegraphics[width=0.95\columnwidth,trim={0 0 0 40},clip]{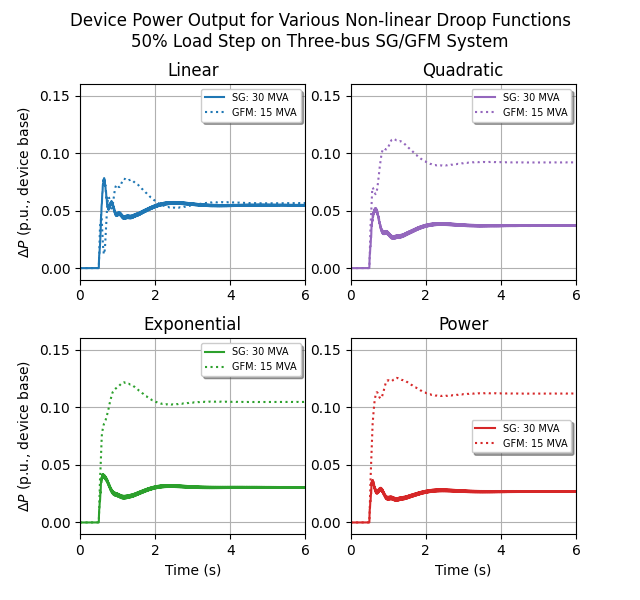}
    \caption{Power output of the SG and GFM inverter for various non-linear droop gains.}
    \label{fig: Power Nonlinear}
\end{figure}

The per unit power deviations of each device, for each of the four functions, is shown in Fig. \ref{fig: Power Nonlinear}. It is clear that each of the non-linear functions permit a larger delivery of active power to the network than the linear function result. The exponential function shows the smallest component of high frequency oscillations immediately after the disturbance. These test results indicate the potential for various non-linear functions to be used for GFM control in future power systems. Clearly, power is not shared equitably between the devices after this initial primary frequency response, but this deviation would be ameliorated by the autonomous power sharing controller first presented in \cite{kenyon_droop-e_2022}. The application of this controller is beyond the scope of this work.

%% file: sections/5_numerical_network.tex
\FloatBarrier
\subsection{Network-Level Dynamics}
\label{sec:numerical: network}

Now, the network level numerical analysis results from simulations with the the IEEE 9- and 39- bus test systems in the EMT domain are presented. 

\subsubsection{Test Case I: IEEE 9-Bus System}
\label{sec: nine bus system}

Simulations on the IEEE 9-bus test system \cite{manitoba_hydro_international_ltd_ieee_nodate-1,kenyon_pypscad_2020}, consisted of a 0.1 pu load step (31.5 MW, 11.5 MVar) at bus 6. Interconnection scenarios of SGs and GFMs are created by systematically supplanting an SG with a GFM, as shown in Table \ref{tab:9 Bus Results}. 

\begin{table}[htbp]
    \small
    \centering
    \caption{9-bus Configuration and Results}
    \setlength\tabcolsep{3pt}    
    \begin{tabular}{c|c|c|c|c|c|c|c}
    \hspace{0.1cm} & \multicolumn{3}{c|}{Device at Bus} & Inertia &ROCOF& Nadir & Damping\\
    Scenario & \multicolumn{1}{c}{1} & \multicolumn{1}{c}{2}&3&(s)& (Hz/s)&(Hz)& ($\zeta$)\\\hline\hline
    A & SG & SG & SG & 4.0 & 0.50 & 59.72 & 0.363\\
    B & GFM & SG & SG & 2.6 & 0.73 & 59.76 & 0.528\\
    C & GFM & GFM & SG & 1.3 & 1.12 & 59.79& 0.842 \\
    D & GFM & GFM & GFM & 0.0 & 1.61 & 59.83 & 0.919
    \end{tabular}
    \label{tab:9 Bus Results}
\end{table}

\begin{figure}[htbp]
    \centering
    \includegraphics[width=0.8\columnwidth,trim={0 0 0 40},clip]{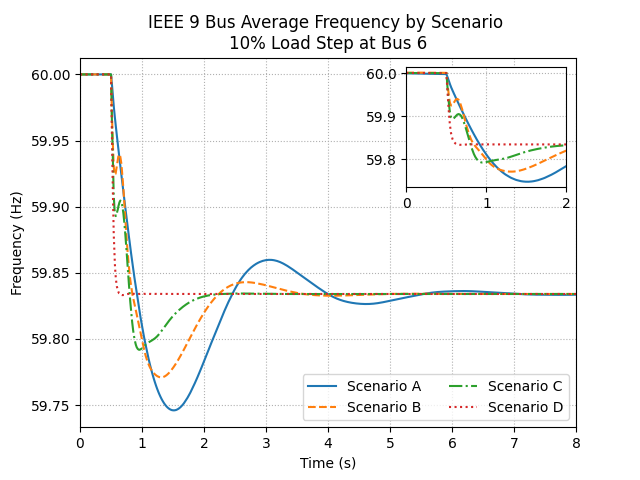}
    \caption{IEEE 9-bus average system frequency response for varied quantities of GFM/SGs. Although the peak ROCOF grows with fewer online SGs, the nadir is simultaneously reduced.}
    \label{fig:9 bus varied inertia}
\end{figure}

\begin{figure}[htbp]
    \centering
    \includegraphics[width=0.8\columnwidth,trim={0 0 0 35},clip]{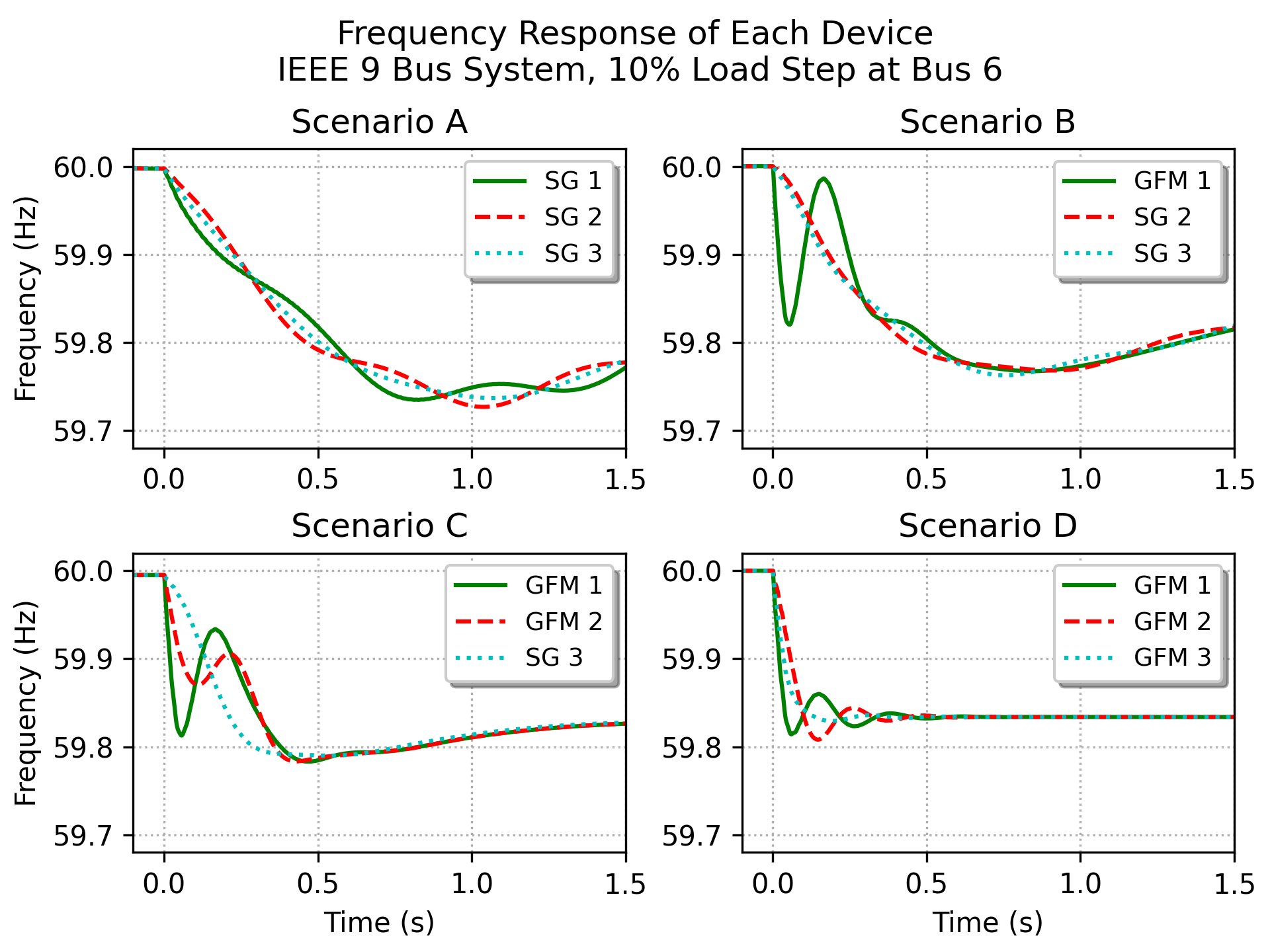}
    \caption{Initial frequency response of each device for the four simulated scenarios on the 9-bus system. Note the contrary motion present with mixed systems (scenario B and C), indicating an initial lack of broad synchronization.}
    \label{fig:9 bus device frequencies}
\end{figure}

Figure \ref{fig:9 bus varied inertia} shows the results for four simulations (scenarios A, B, C, and D - explained in Table \ref{tab:9 Bus Results}) where the SGs are systematically replaced by the GFMs. The reduced values of inertia are captured in Table \ref{tab:9 Bus Results}, along with the resultant nadir and ROCOF. It is evident that as the mechanical inertia is decreased, the nadir is raised while the ROCOF increases, which is indicative of the dominant first order response of the additional GFMs as the inverse would be expected in a second order system. The system frequency oscillation period with all SGs (Scenario A) matches the single machine step response; i.e., 0.4 Hz oscillations. By inspection of Fig. \ref{fig:9 bus device frequencies}, the dominant oscillatory mode increases in frequency with more GFMs. The damping ratio from Table \ref{tab:9 Bus Results} shows a clear increase as the converters are changed to GFMs.

Individual device frequencies for each Scenario are presented in Fig. \ref{fig:9 bus device frequencies} For Scenario A, all three SGs have similar frequency trajectories; the three devices maintain broad synchronization following the perturbation. Herein lies the motivation for the center of inertia and average frequency metrics, which rely on the assumption of similar (i.e., second order) frequency trajectories of all devices \cite{sauer_power_2017,lasseter_grid-forming_2020}.

The Scenario B system frequency is second order with a damping value around 50\%; the dominant mode damping is increased according to Table \ref{tab:9 Bus Results}. From Fig. \ref{fig:9 bus device frequencies}, it is obvious that the three devices are not broadly synchronized immediately following the disturbance. Scenario C frequency exhibits overshoot, but a quasi-first order recovery; i.e. the concavity in the green trace from t = 1.5--2.5 s is the inverse of the expected  second order recovery. The GFM device frequencies change too rapidly for the devices to remain loosely synchronized during the first 0.5 s following the perturbation. With all GFM devices (D), the system frequency follows a first order response with overshoot and frequency oscillations absent. 

\begin{figure}[htbp]
    \centering
    \includegraphics[width=0.8\columnwidth,trim={0 0 0 40},clip]{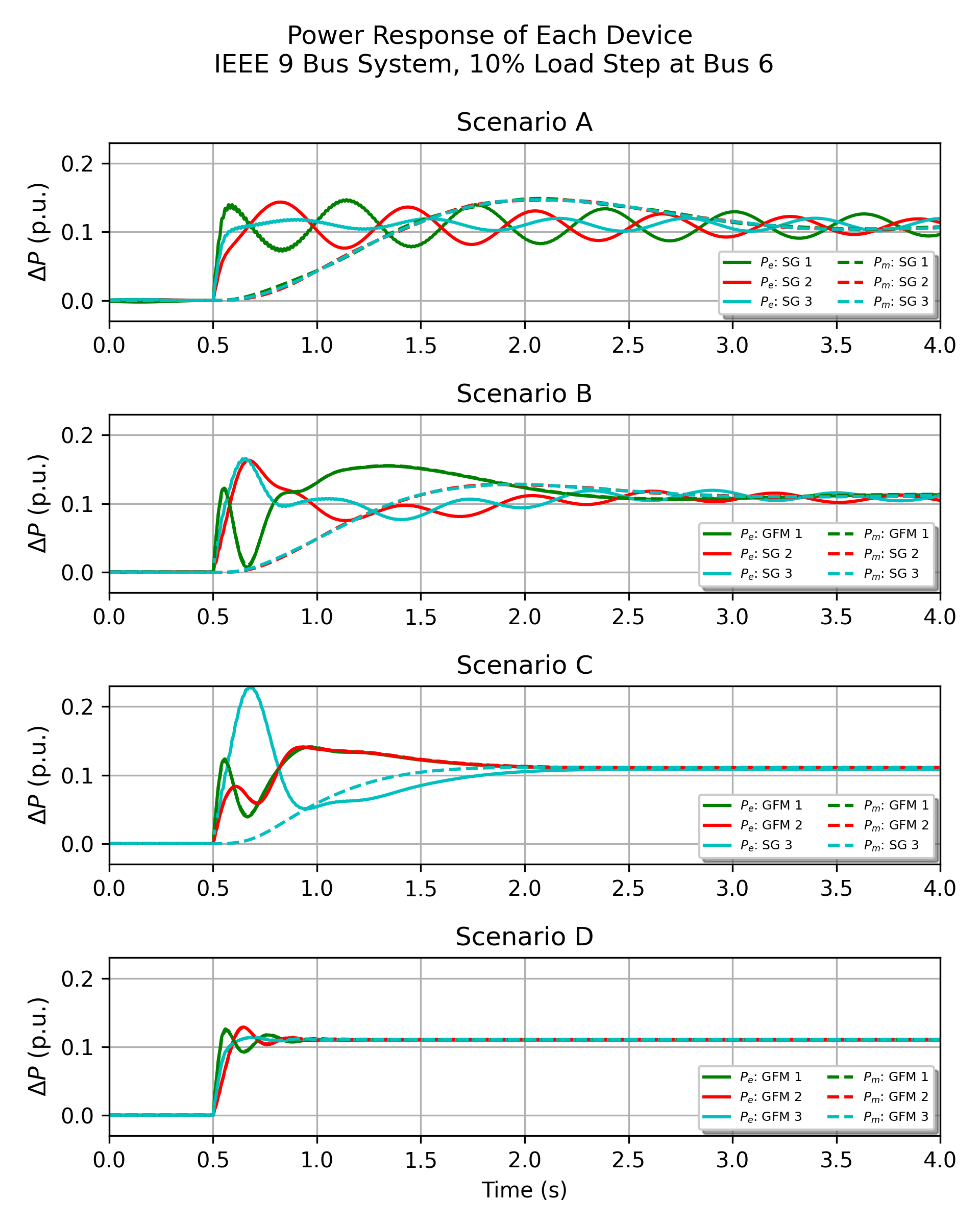}
    \caption{Comparison of pre-converter and electrical powers of each device for the four 9-bus simulation scenarios. 
    }
    \label{fig:9 bus powers}
\end{figure}

The $p_e$ and $p_m$ response for each scenario are presented in Fig. \ref{fig:9 bus powers}. The electrical powers show inter-area oscillations ($f = 1.6$Hz) between SG 1 and SG 2 in Scenario A. Note the time separation between these $p_{e,G}$ oscillations and $p_{m,G}$. Scenario B shows the very rapid changes in $p_{m,I}$--$p_{e,I}$ (these are indistinguishable at this temporal resolution) of GFM 1, which exacerbates a larger peak $p_{e,G}$ of SG 2 and 3, while the subsequent oscillations are more damped. The peak $p_{e,G}$ output of SG 3 in Scenario C is further increased, although there are no oscillations following this overshoot. The conclusion is that the rapid frequency changes of the GFMs, evident in Fig. \ref{fig:9 bus device frequencies}, cause the network conditions to change rapidly, while the SG frequency follows the slower second order response resulting in a larger $p_{e,G}$ extraction; the fast frequency change of the GFM forces the SG into larger oscillations because of the slower frequency change, which exacerbates the dynamic excursions. This large $p_{e,G}$ excursion corroborates the hypothesis of an increasing initial extraction of energy when the SG rating ratio is low with high shares of GFM, based on the observations in the two generator system. Scenario D power outputs show a reduction in power oscillations with minimal overshoot and a rapid arrival to settling outputs. 

The results from the 9-bus system suggest that the presence of GFM inverters reduces the average system frequency nadir, while increasing ROCOF. Additionally, dominant oscillatory mode analysis indicates a monotonic increase in damping with more GFMs. 

\FloatBarrier
\subsubsection{Test Case II: IEEE 39-Bus System}
\label{sec: 39 bus}

The IEEE 39-bus test system \cite{manitoba_hydro_international_ltd_ieee_nodate} is presented as a larger case study. Supporting Python code and details about this model are  available open-source at \cite{kenyon_pypscad_2020}. All ten initial SG devices are systematically replaced by GFMs, with a scenario defining each iteration; i.e. scenarios 0--10 in Table \ref{tab:39 Bus Results}. The 0.1 pu load step (600 MW/140 Mvar) occurs at bus 15.

\begin{table}[htbp]
    \small
    \centering
    \caption{39-bus Configuration and Results}
    \setlength\tabcolsep{1pt}   
    \begin{tabular}{c|c|c|c|c|c|c}
    \multirow{2}{*}{\rotatebox{0}{Scenario}} & GFMs & Inertia & ROCOF & Nadir & Damping & Dom. Mode\\
     & at Buses & (s) & (Hz/s) & (Hz) & ($\zeta$) & Freq. (Hz)\\\hline\hline
    0 & n/a & 4.0 & 0.567 & 59.690 & 0.361 & 0.324\\
    1 & 30 & 3.6 & 0.587 & 59.712 & 0.394 & 0.332\\
    2 & 30--31 & 3.2 & 0.669 & 59.717 & 0.440 & 0.345\\
    3 & 30--32 & 2.8 & 0.808 & 59.724 & 0.489 & 0.358\\
    4 & 30--33 & 2.4 & 0.930 & 59.730 & 0.551 & 0.369\\
    5 & 30--34 & 2.0 & 1.071 & 59.738 & 0.622 & 0.383\\
    6 & 30--35 & 1.6 & 1.225 & 59.748 & 0.722 & 0.386\\
    7 & 30--36 & 1.2 & 1.396 & 59.748 & 0.958 & 0.215\\
    8 & 30--37 & 0.8 & 1.525 & 59.756 & 0.382 & 1.320\\
    9 & 30--38 & 0.4 & 1.648 & 59.772 & 0.224 & 1.414\\
    10 & All GFM & 0.0 & 1.852 & 59.808 & 0.893 & 1.99
    \end{tabular}
    \label{tab:39 Bus Results}
    \vspace{-1em}
\end{table}

\begin{figure}[htbp]
    \centering
    \includegraphics[width=0.8\columnwidth,trim={0 0 0 40},clip]{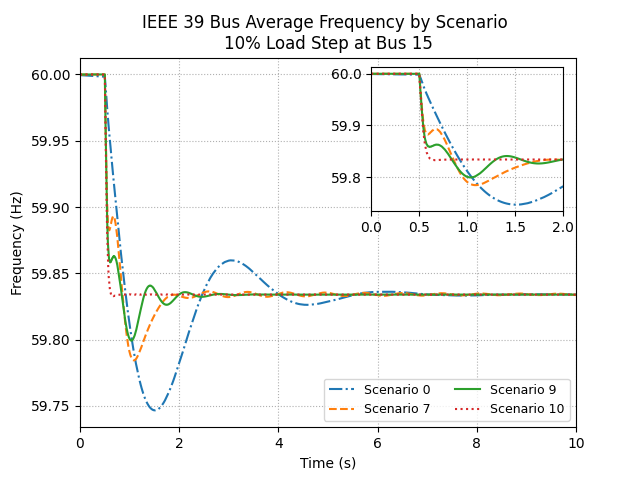}
    \caption{Average frequency response of 39- bus system for varied quantities of GFMs and SGs and a 0.1 pu load step at bus 15.}
    \label{fig:39 bus ave freq}
    \vspace{-1em}
\end{figure}

Figure \ref{fig:39 bus ave freq} shows the average frequency for each of the 11 scenarios simulated on the 39-bus system; there are broad similarities to the 9-bus results, with a substantial reduction in overshoot following the transition to a GFM dominated system. The inverted period within the initial decceleration occurs at higher average frequency values with a larger quantity of GFM devices. The frequency statistics presented in Table \ref{tab:39 Bus Results} show that while ROCOF increases with larger quantities of GFMs and a resultant decrease in system mechanical inertia, the nadir is simultaneously raised; the average frequency trajectories expose a transition to a first order response. 

\begin{figure}[htbp]
\vspace{-1em}
    \centering
    \includegraphics[width=0.8\columnwidth,trim={0 0 0 40},clip]{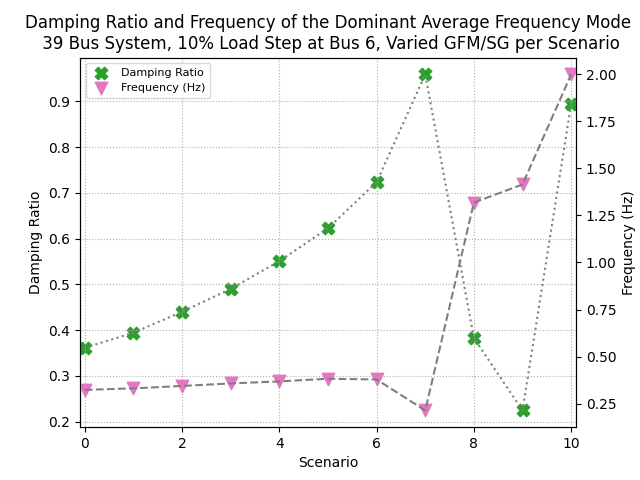}
    \caption{Damping ratio and frequency of the dominant average frequency mode for the 39-bus simulations.}
    \label{fig: 39 Damping Modes}
    \vspace{-1em}    
\end{figure}

\begin{figure}[htbp]
    \centering
\vspace{-1em}    
    \includegraphics[width=0.8\columnwidth,trim={0 0 0 40},clip]{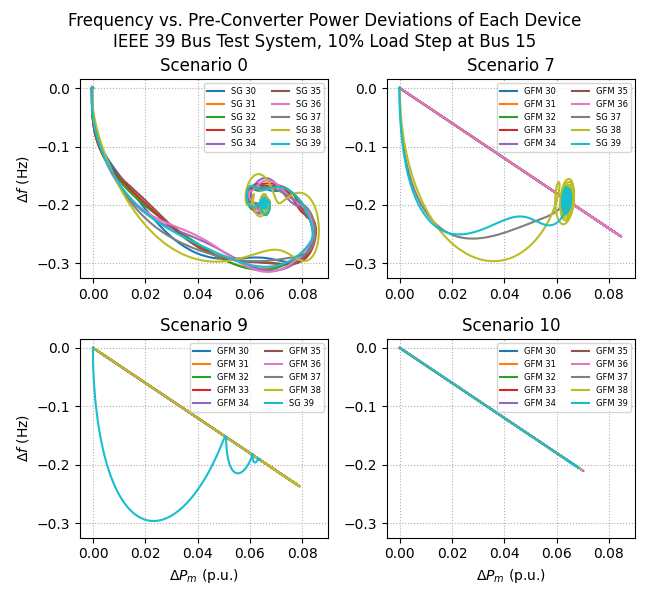}
    \caption{Frequency--power portraits of each device for scenario 0, 7, 9, and 10 for the 10\% load step on the 39-bus test system.}
    \label{fig: 39 Pf portraits} 
\end{figure}

Also presented in Table \ref{tab:39 Bus Results} is the damping and frequency of the primary mode of the average frequency for each scenario. These two sets of data are reproduced graphically in Fig. \ref{fig: 39 Damping Modes}. It is apparent that as the first half of converters are swapped from SGs to GFMs, there is a large increase in damping of the primary mode, while the frequency of the primary mode sees a small increase. The monotonicity of these progressions degrades beyond Scenario 7, where a substantial increase in mode frequency is present with greatly varying levels of damping. In fact, contrary to the steady increase in damping seen in the 9-bus system simulations, a substantial decrease in damping can be seen with only a few SGs remaining, though at a greatly increased dominant mode frequency. From Fig. \ref{fig: 39 Pf portraits}, Scenario 9, a larger number of distinct oscillations are seen in the remaining SG. The presence of a large number of first order devices is exacerbating the second order response of the remaining SGs. The results here indicate that neither the network, nor the quantity of the devices, plays a role in these high frequency modes present during times of low SG rating ratios. 

Portraits of the $f$--$p_{m,I}$ trajectories for a selection of 39-bus system simulations are presented in Fig. \ref{fig: 39 Pf portraits}, where the subtitles correspond with the Table \ref{tab:39 Bus Results} entry. With no GFMs, the SGs follow the trajectory of an initial frequency deviation prior to pre-converter changes. Prior to convergence on the steady state values, the trajectories exhibit oscillations in the form of converging spirals. With half GFMs in scenario 5, the algebraic relation between frequency and pre-converter power is evident, while the SG trajectories are shortened with less overshoot. With only a single SG online in scenario 9, the SG exhibits no pre-converter power overshoot; however, substantial frequency oscillations are present, of a higher frequency as confirmed by the change in primary mode frequency. 

%% file: sections/6_maui.tex
\section{Realistic Power System: Maui Island}
\label{sec: maui system}

\begin{figure}[htp!]
    \centering
    \fbox{\includegraphics[width=0.95\columnwidth,trim={2 2 2 2},clip]{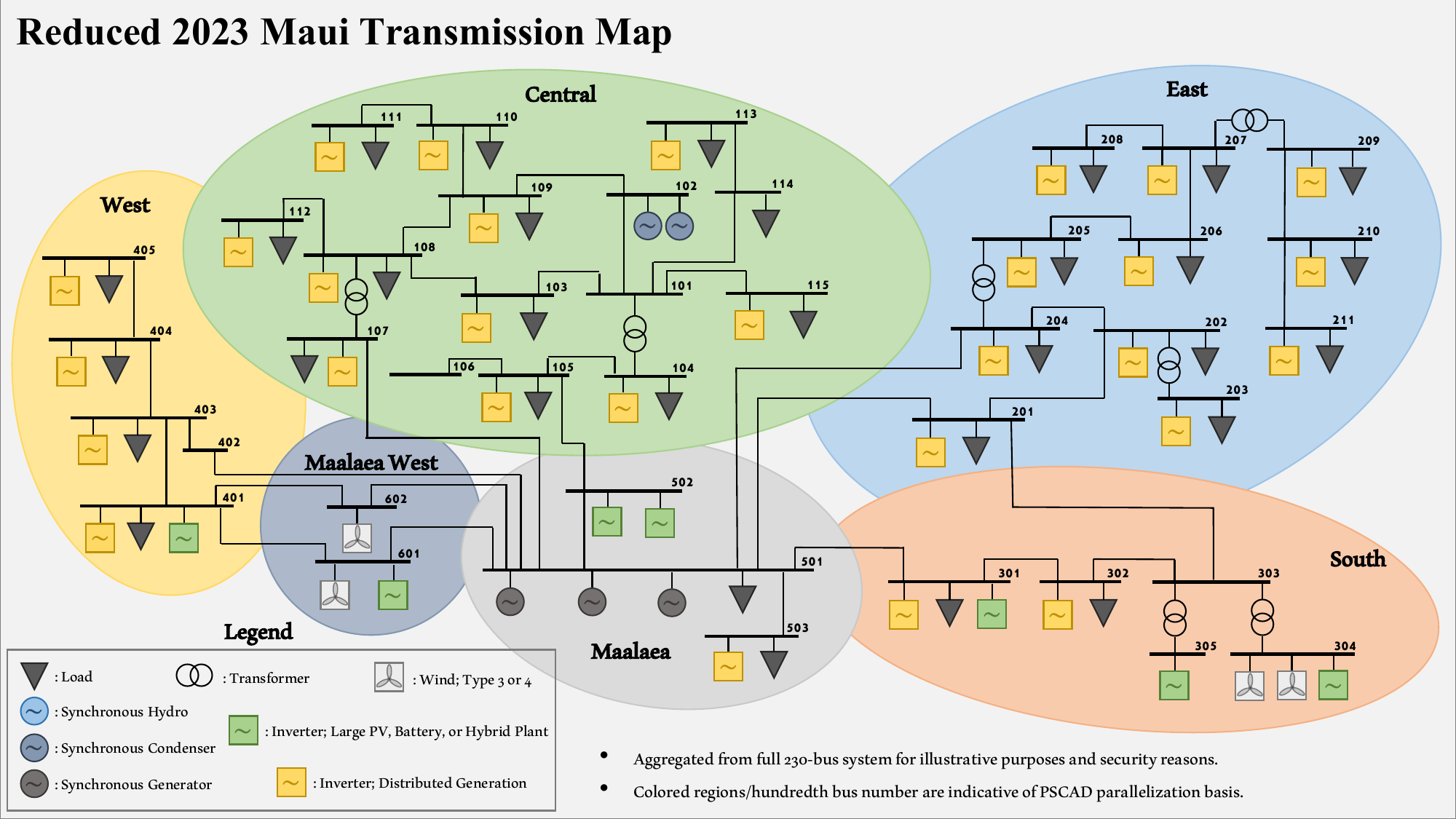}}
    \caption{Reduced representation of the Maui power system: adopted from \cite{kenyon_criticality_2023}}
    \label{fig: maui system}
\end{figure}

In this section, a validated power system model of the Hawaiian Island of Maui is simulated to identify the transition trajectory to a first order frequency response with a GFM device in a realistic system. The Maui power system (Fig. \ref{fig: maui system}) is a 200 MW peak demand, 69 kV mesh network that has a variety of generation resources including synchronous generators, synchronous condensers, type 3 and 4 wind plants, hybrid power plants (HPPs), battery energy storage systems (BESSs), utility solar plants, and distributed generation (DG) in excess of 50\% of demand at peak output. A model of this power system has been developed in PSCAD, validated against field data, and parallelized on 32 cores to make simulations computationally tractable \cite{kenyon_validation_2021}; it is one of the largest, validated EMT domain power system models to date \cite{hoke_island_2021}. Even for this relatively small system, a single 20 second simulation requires over four hours to run on a high powered computer, highlighting the challenges of EMT domain simulations on large power systems. This section compares the frequency response of the Maui power system when a single PV-BESS hybrid power plant (HPP) is converted from GFL to GFM. 

The dispatch of the system is highlighted in Table \ref{tab:System Attributes}, where the regions correspond to the reduced transmission single line diagram in Fig. \ref{fig: maui system}. With IBR penetration defined as $\eta = P_{IBR}/P_{Tot}$, the current dispatch is 96\% IBRs with $P_{IBR} = 140.2 MW$ and $P_{Tot} = 144.7 MW$. With a large number of synchronous condensers online, the system equivalent inertia of the dispatch is $H = 0.97s$. There are two 30 MVA HPPs colocated at bus 502 in the Maalaea region, which are configured as GFL devices with frequency-droop (4\%) grid support; they both are dispatched at 2.85 MW. The HPP in the South region is not altered. Two simulations are executed to depict the impacts of a GFM device on the Maui network. Both simulations have the same dispatch as Table \ref{tab:System Attributes}. The first simulation has all three HPPs as GFL devices with grid support, which is referred to as the \textit{base-case}. The second simulation has a single HPP at bus 502 converted to GFM control; this case is referred to as the \textit{GFM-case}. The perturbation on the system is the disconnection of both type 4 wind plants, which share a common point of interconnection, for a loss of 21 MW, equating to 15\% of total generation.

\input{sections/attribute_table}

\begin{figure}[!htbp]
    \centering
    \includegraphics[width=0.8\columnwidth,trim={0 0 0 40},clip]{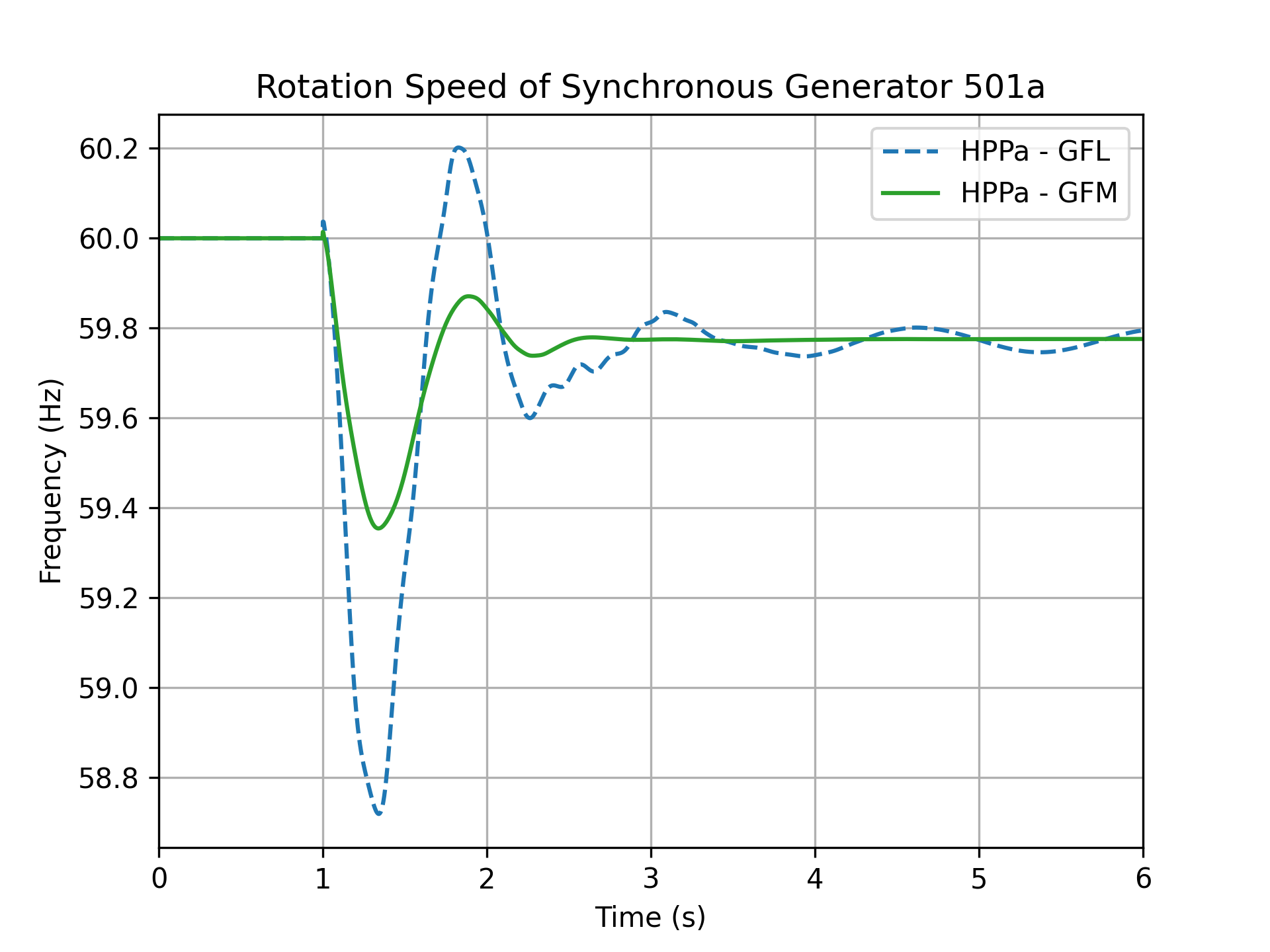}
    \caption{Frequency response of SG 501a for the base-case and GFM-case simulations.}
    \label{fig: maui freq}
\end{figure}

The rotational speed of SG 501a is scaled to Hz, and presented as a proxy for the frequency response of the system for both simulations in Fig. \ref{fig: maui freq}. The ROCOF and nadir for the base-case are 6.7 Hz/s and 58.7 Hz, respectively. In the GFM-case, there is a clear improvement in the frequency response, evidenced by improved damping of the SG response, a less deviant nadir of 59.35 Hz, and a ROCOF of 3.7 Hz/s. As expected from the previous derivations and simulation results, the frequency response with a single GFM device online shows the trend of a transition towards a first order response. 

\begin{figure}[!htbp]
    \centering
    \includegraphics[width=0.8\columnwidth,trim={0 0 0 40},clip]{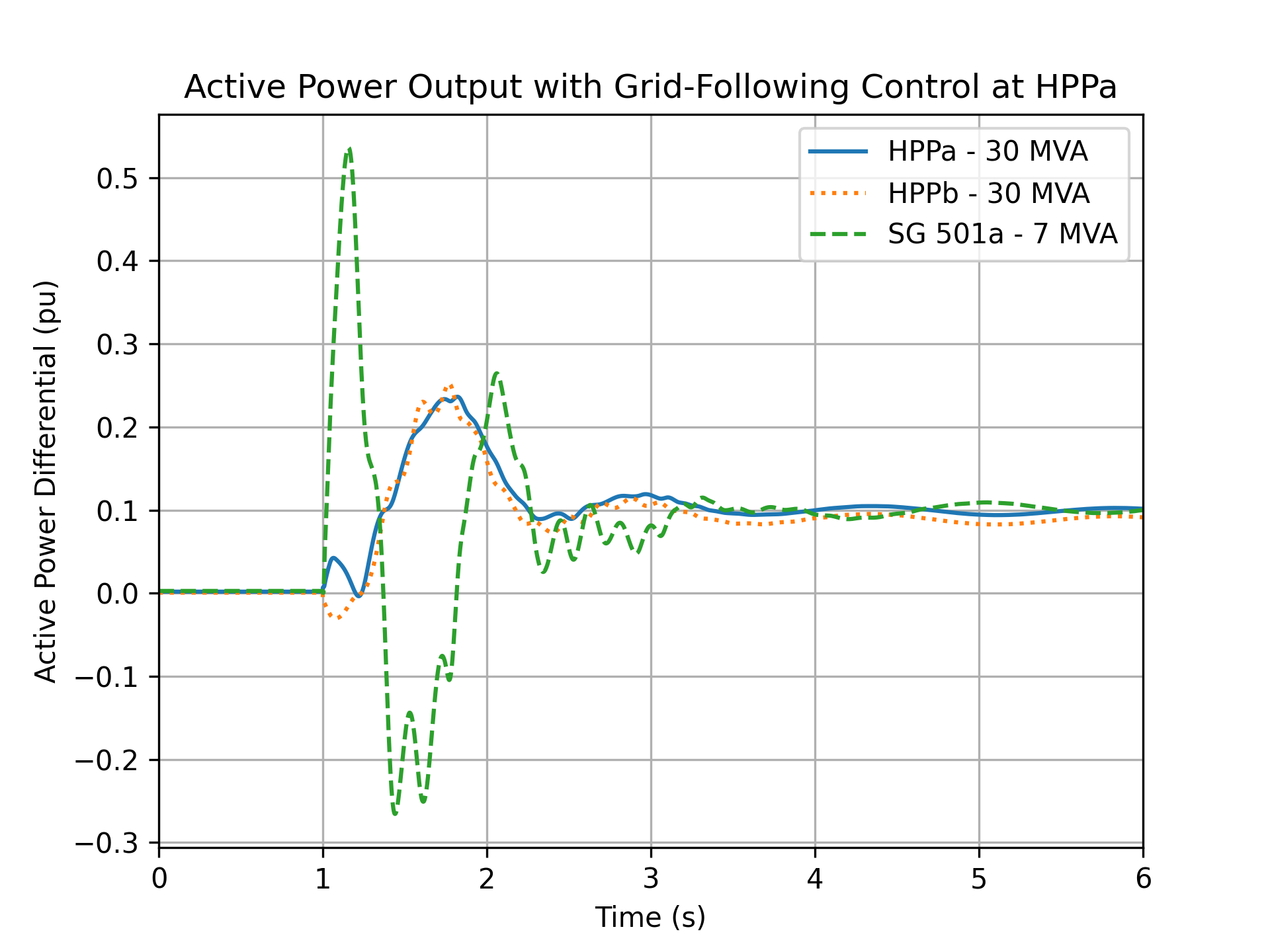}
    \caption{Active power response of two HPPs and SG 501a for the base-case.}
    \label{fig: maui GFL}
\end{figure}

\begin{figure}[!htbp]
    \centering
    \includegraphics[width=0.8\columnwidth,trim={0 0 0 40},clip]{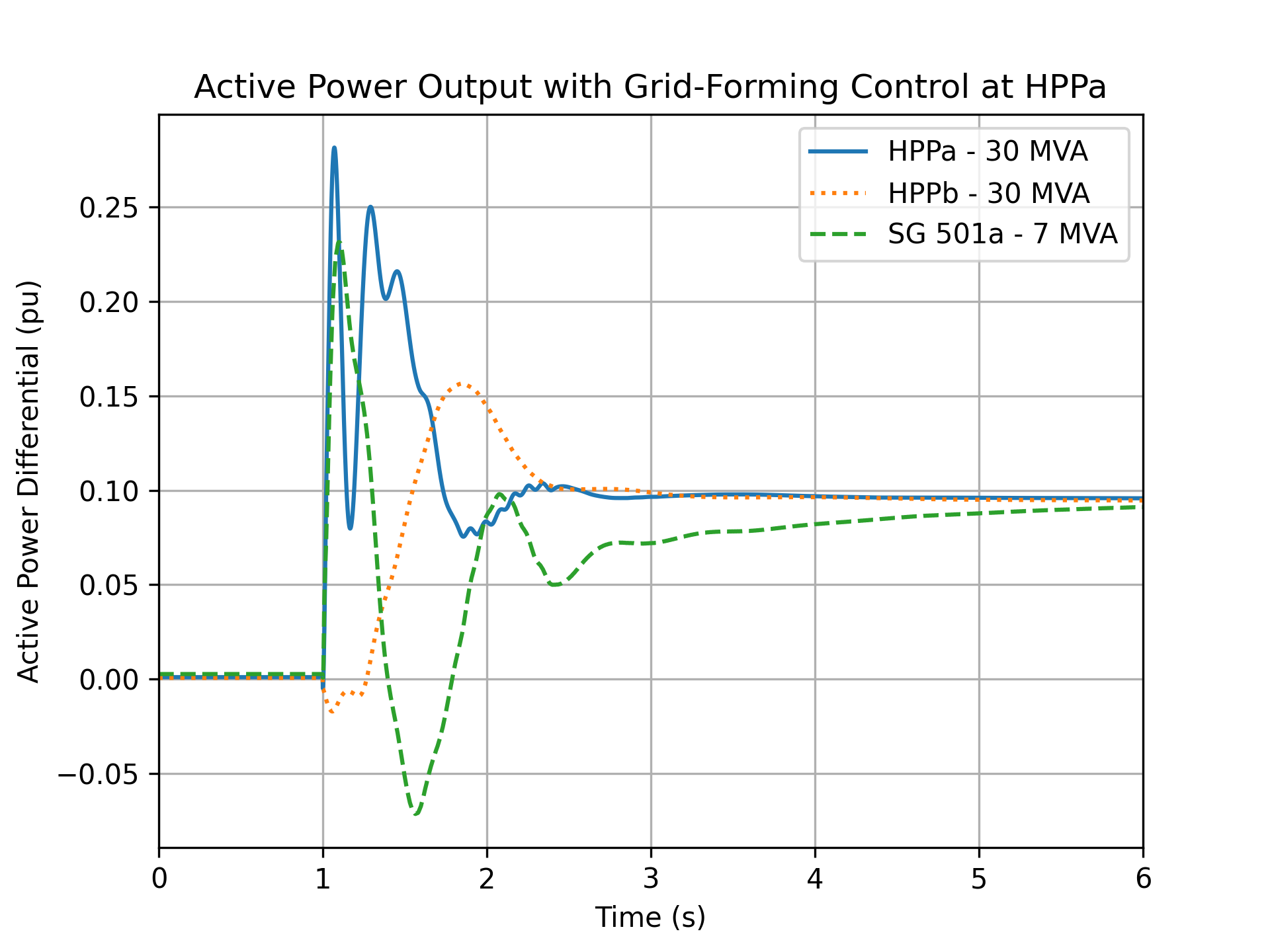}
    \caption{Active power response of two HPPs and SG 501a for the GFM-case.}
    \label{fig: maui GFM}
\end{figure}

The active power response of SG 501a and the two HPP units is presented for each simulation in Figs. \ref{fig: maui GFL} and \ref{fig: maui GFM}. In the base-case, the responses of the HPPs are nearly identical as both devices are operating under a frequency droop grid-supporting control strategy. There are high frequency oscillations present, primarily in the SG response. With HPPa transitioned to a droop controlled GFM, the immediate (<0.5s) response of the SG and GFM HPP are similar, because both devices are regulating the local voltage, with the result a power extraction from the network as opposed to controller driven power injections. The peak power extraction from the SG is reduced by 50\%, from 0.5 pu in the base-case, to 0.25 pu in the GFM-case. Improved damping of the high frequency mode of the SG present in Fig. \ref{fig: maui GFL}, and the low frequency mode of the frequency response in Fig. \ref{fig: maui freq} is evident in the active power output traces.


The mode switch of a single HPP on the validated Maui model from GFL to GFM during a period of very high instantaneous shares of IBRs operating as GFLs is highly beneficial to the frequency stability of the system. The impacts of a very large generation loss are mitigated with the single GFM, as the system frequency shows damping on multiple oscillatory modes, while the frequency response displays the trend towards a first order system as previously discussed and demonstrated on the test systems.

%% file: sections/attribute_table.tex
\begin{table}[!hbt]
    \centering
    \caption{Summary of Maui Region Resources For Studied Dispatch; distributed generation (DG), battery energy storage system (BESS), hybrid power plant (HPP).}
    {\renewcommand{\arraystretch}{0.8}
    \small\setlength{\tabcolsep}{.4em}
    \begin{tabular}{c|ccc}
      \multirow{2}{*}{\rotatebox{0}{\textbf{Region}}} & \multirow{2}{*}{\textbf{Type}} & \textbf{Rating} & \textbf{Output}\\
       & & (MVA) & (MW/Mvar) \\\addlinespace[2pt]\hline\hline\addlinespace[2pt]
      \multirow{2}{*}{\rotatebox{0}{Central}} & \multicolumn{1}{l}{\tabitem 2x Condensers} & 29.1 & 0.0 / 0.0\\
      & \multicolumn{1}{l}{\tabitem 4x Hydro} & 6.4 & 0.0 / 0.0\\
      & \multicolumn{1}{l}{\tabitem 66x DG} & 45.2 & 41.1 / -3.1 \\\addlinespace[2pt]\hline\addlinespace[2pt]
      \multirow{1}{*}{\rotatebox{0}{East}} & \multicolumn{1}{l}{\tabitem 33x DG} & 13.1 & 12.1 / -1.3
      \\\addlinespace[2pt]\hline\addlinespace[2pt]
      \multirow{5}{*}{\rotatebox{0}{South}} & \multicolumn{1}{l}{\tabitem 2x Type 4 Wind} & 24.0 & 21.0 / 0.0 \\
      & \multicolumn{1}{l}{\tabitem 1x BESS} & 10.0 & 0.0 / 0.0 \\
      & \multicolumn{1}{l}{\tabitem 1x Utility Solar} & 2.9 & 2.7 / 0.0 \\
      & \multicolumn{1}{l}{\tabitem 1x HPP} & 19.1 & 0.0 / 0.5 \\
      & \multicolumn{1}{l}{\tabitem 27x DG} & 33.5 & 30.5 / -2.2 \\\addlinespace[2pt]\hline\addlinespace[2pt]
      \multirow{2}{*}{\rotatebox{0}{West}} & \multicolumn{1}{l}{\tabitem 1x Utility Solar} & 2.9 & 2.7 / 0.0 \\
      & \multicolumn{1}{l}{\tabitem 42x DG} & 19.1 & 17.4 / -1.8  \\\addlinespace[2pt]\hline\addlinespace[2pt]
      \multirow{4}{*}{\rotatebox{0}{Maalaea}} & \multicolumn{1}{l}{\tabitem 3x Generators} & 21.0 & 5.7 / 6.9\\ 
      & \multicolumn{1}{l}{\tabitem 4x Condensers} & 107.2 & 0.0 / 14.9\\
      & \multicolumn{1}{l}{\tabitem 2x HPP} & 60.0 & 5.7 / 5.3\\
      & \multicolumn{1}{l}{\tabitem 3x DG} & 3.5 & 3.2 / -0.3
      \\\addlinespace[2pt]\hline\addlinespace[2pt]
      \multirow{1}{*}{\rotatebox{0}{Maalaea}} & \multicolumn{1}{l}{\tabitem 2x Type 3 Wind} & 56.7 & 3.9 / 0.0 \\
      \multirow{1}{*}{\rotatebox{0}{West}} & \multicolumn{1}{l}{\tabitem 1x BESS} & 10.0 & 0.0 / 0.0
      \\\addlinespace[2pt]\hline\addlinespace[2pt]
      \multirow{3}{*}{\textbf{Totals}} & \hspace{0.3cm} \textit{Synchronous:} 13 & 163.7 & 5.7 / 21.8\\
      & \hspace {0.3cm} \textit{Inverter Based:} 11 & 205.5 & 35.9 / 5.8\\
      & \textit{DG:} 171 & 114.8 & 104.3 / -8.7 
    \end{tabular}}
    \label{tab:System Attributes}
\end{table}

%% file: sections/7_discussion.tex
\section{Discussion}
\label{sec:discussion}
For devices that adjust pre-converter power as a function of frequency changes such as SGs, a larger ROCOF generally yields a deeper nadir for the same magnitude power imbalance\cite{inverter-based_resource_performance_task_force_fast_2020}. Consider the approximating equation $\label{eq: inertial}\Delta f_{prior} = \alpha_{ROCOF}  \times t_{response}$ where $\Delta f_{prior}$ is the frequency deviation prior to substantive $p_{m,G}$ changes, $\alpha_{ROCOF}$ is the ROCOF for a generic system, and $t_{response}$ is the pre-converter power response time (as used in \eqref{eq:GFM power} and \eqref{eq:SG power}). A relatively larger ROCOF for the same $t_{response}$ (such as $\tau_G$, in \eqref{eq:SG power}) yields a larger $\Delta f_{prior}$ prior to substantial $p_m$ changes. This is the inertial response period when rotational kinetic energy ($E_{int,G}$) is extracted from the SGs; in SG dominated systems, less inertia yields larger $\alpha_{ROCOF}$ values, increasing susceptibility to lower nadirs that can trigger frequency load shedding \cite{north_american_electric_reliability_corporation_standard_2017}. This behavior has been highlighted as exacerbated by inverter dominated systems \cite{ulbig_impact_2014}.

Due to the control design, the GFM relation is $\Delta f_{prior} = 0$; after $p_{e,I}$ changes due to varying network conditions, $p_{m,I}$ is matched prior to any change in frequency. The GFM changes frequency according to a control objective, not rotational kinematics and frequency-triggered governor action to match $p_{e,G}$ and $p_{m,G}$. Given this flexibility, the pre-converter--frequency relationship can be of a lower order with GFMs, and potentially redefine the frequency response in GFM dominated systems. Here, it is noted that there are potential broader issues due to larger ROCOFs such as relay tripping, device disconnection, and machine shaft strain. While the analysis of these considerations is beyond the scope of this work, it is noted that the non-linear droop functions presented in Section \ref{sec:potential nonlinear}, which maintain a first order response, reduced the ROCOF in all cases. 

An unusual phenomenon was uncovered in the two generator system, where for very low levels of SGs as a rating ratio (i.e., when the rating of the SG was $<$ 20\% of the total system online generation), a previously unreported high frequency oscillatory mode was observed. This mode was uncovered in the two generator system and also observed in the 39-bus system with ten generators, where similar rating ratios could be realized by binary dispatch changes where a generator is either a GFM or SG. That the mode exists in both of these systems indicates that it is not a function of generator quantity nor network interactions; it is likely the result of the rapid frequency change of GFMs, which places a substantial $p_{e,G}$ differential on the small amount of inertia remaining on the system. Interestingly, while it is caused by the rapid change in frequency of the GFMs, the GFMs are also the source of damping in the system as no substantial $p_{m,G}$ changes are observed prior to the oscillatory mode dampening out. This observation is generally contradictory to the claim of a trend of increased damping with GFM devices \cite{lasseter_grid-forming_2020,sajadi_synchronization_2022}, in that there is evidently an inversion in this damping when GFMs are the dominant resource in a heterogeneous system. This oscillatory mode was not observed in the 9-bus system because the discrete variation of generators did not yield this lower rating ratio ($<20\%$).

%% file: sections/8_conclusion.tex
\FloatBarrier

\section{Conclusion}
\label{sec:conclusion}

This paper investigated the active power--frequency conversion dynamics of the synchronous generator and the droop controlled grid-forming inverter and the associated power system frequency impacts due to the device disparities. Analysis highlighted the lower order frequency dynamics and the inverted frequency regulation of grid-forming inverters, as compared to synchronous generators, which laid the framework for the introduction of novel, non-linear frequency droop control methods. Electro-magnetic transient simulations at the device level with full order models confirmed this power--frequency order reduction, while investigation into the DC-side dynamics concluded on a negligible impact with proper design, which supports the approximation of an immediate through-put active power capability of grid-forming devices. Simulations with the non-linear frequency droop functions yielded improved frequency dynamics with less deviant nadirs and smaller rates of change of frequency, encouraging further work in the non-linear droop control field. Network-level electromagnetic transient simulations of the IEEE 9- and 39- bus systems displayed a general increase in primary frequency mode damping with a transition to dominating levels of grid-forming inverters, while at very low levels of inertia, a high frequency mode was uncovered. Simulations with a validated model of the Hawaiian island of Maui power system concluded the study with results indicating a transition towards first order frequency dynamics with grid-forming inverters.